\documentclass[aps,amsmath,amssymb,11pt,tightenlines,superscriptaddress,preprintnumbers,nofootinbib]{revtex4}
\pdfoutput=1
\usepackage{latexsym}
\usepackage{amsthm}
\usepackage{float}
\usepackage{amsmath,amssymb}
\usepackage{graphicx,subfigure,grffile}
\usepackage{hangcaption,fancybox,feynmp}
\usepackage{indentfirst}
\usepackage{epsfig,subfigure,psfrag}
\usepackage{dcolumn}
\usepackage{bm}
\usepackage{slashed}
\usepackage{cancel}
\usepackage{color}
\usepackage{tabularx}
\usepackage[titletoc]{appendix}
\usepackage{hyperref}
\usepackage[usenames,dvipsnames]{xcolor}
\usepackage[normalem]{ulem}

\preprint{MITP/15-036, IFT-UAM/CSIC-15-047, FTUAM-15-13}

\def\slashchar#1{\setbox0=\hbox{$#1$}           
   \dimen0=\wd0                                 
   \setbox1=\hbox{/} \dimen1=\wd1               
   \ifdim\dimen0>\dimen1                        
      \rlap{\hbox to \dimen0{\hfil/\hfil}}      
      #1                                        
   \else                                        
      \rlap{\hbox to \dimen1{\hfil$#1$\hfil}}   
      /                                         
   \fi}                                         %

\newcommand{\be}{\begin{eqnarray}}
\newcommand{\ee}{\end{eqnarray}}
\newcommand{\bea}{\begin{align}}
\newcommand{\eea}{\end{align}}

\newcommand{\GeV}{{~\rm GeV}}

\newcommand{\pd}{\partial}

\newcommand{\ev}[1]{\ensuremath{\left\langle #1 %
                     \right\rangle}} 

\newcommand{\BR}{\text{BR}}

\newcommand{\heavymed}{{Z^\prime}}
\newcommand{\darkmed}{{A^\prime}}

\begin{document}

\title{Lepton Jets from Radiating Dark Matter}

\author{Malte Buschmann}  \email{buschmann@uni-mainz.de}
\author{Joachim Kopp}  \email{jkopp@uni-mainz.de}
\author{Jia Liu}  \email{liuj@uni-mainz.de}
\affiliation{PRISMA Cluster of Excellence and Mainz Institute for
Theoretical Physics, Johannes Gutenberg University, 55099 Mainz, Germany}

\author{Pedro A.N. Machado}  \email{pedro.machado@uam.es}
\affiliation{
Departamento de F\'isica Te\'orica and Instituto de F\'{\i}sica Te\'orica,
IFT-UAM/CSIC,\\
Universidad Aut\'onoma de Madrid, Cantoblanco, 28049, Madrid, Spain}

\date{\today}

\begin{abstract}
The idea that dark matter forms part of a larger dark sector is very
intriguing, given the high degree of complexity of the visible sector.  In this
paper, we discuss lepton jets as a promising signature of an extended dark
sector.  As a simple toy model, we consider an $\mathcal{O}(\text{GeV})$ DM
fermion coupled to a new $U(1)'$ gauge boson (dark photon) with a mass
of order GeV and kinetically mixed with the Standard Model photon.
Dark matter production at the LHC in this model is typically accompanied by
collinear radiation of dark photons whose decay products can form lepton jets.
We analyze the dynamics of collinear dark photon emission both analytically and
numerically. In particular, we derive the dark photon energy spectrum using
recursive analytic expressions, using Monte Carlo simulations in Pythia, and
using an inverse Mellin transform to obtain the spectrum from its moments.  In
the second part of the paper, we simulate the expected lepton jet signatures
from radiating dark matter at the LHC, carefully taking into account the
various dark photon decay modes and allowing for both prompt and displaced
decays.  Using these simulations, we recast two existing ATLAS lepton jet
searches to significantly restrict the parameter space of extended dark sector
models, and we compute the expected sensitivity of future LHC searches.
\end{abstract}

\maketitle


\section{Introduction}
\label{eq:introduction}

For decades, nature has been keeping us completely in the dark about
the nature of dark matter.  In view of this, the scope of theoretical
and phenomenological studies has become significantly broader than it
used to be some years ago.  An often discussed possibility nowadays is
a dark sector featuring new gauge interactions.  Phenomenologically,
such new gauge interactions can for instance lead to Sommerfeld
enhancement of the dark matter (DM) annihilation cross section
\cite{Hisano:2003ec, Cirelli:2008pk, ArkaniHamed:2008qn}, to DM
self-interactions \cite{Spergel:1999mh, Tulin:2013teo}, which may
affect small scale structures in the Universe such as dwarf galaxies,
or to DM bound states~\cite{Shepherd:2009sa, Altmannshofer:2014cla,  Foot:2014uba}. 
In this context, it is worth noting that there are long-standing
discrepancies between theory and small scale structure observations,
which could be explained either by dark matter self interactions
\cite{Spergel:1999mh, Tulin:2013teo} or by baryonic
feedback~\cite{Vogelsberger:2014kha, Sawala:2014xka}. Also recent
observations of the galaxy cluster Abell~3827 could point towards
self-interacting DM~\cite{Massey:2015dkw} (see, however,
\cite{Kahlhoefer:2015vua}).

Here, we discuss the prospects of detecting new gauge interactions in the dark
sector at the LHC. We work in a simplified model with a dark matter fermion
$\chi$ and a $U(1)'$ gauge boson $\darkmed$ (dark photon), kinetically mixed with the
photon \cite{Holdom:1985ag, ArkaniHamed:2008qn}  (see e.g. \cite{Jaeckel:2010ni}
for a review). This model, which we dub \emph{radiating dark matter},
serves here as a proxy for a large class of more complicated scenarios, to which our
conclusions can be applied as well.  If the dark matter and dark photon masses
are relatively small compared to the typical partonic center of mass energy at
the LHC, emission of dark photons from a dark matter particle receives a strong
enhancement in the collinear direction.  This is analogous to the emission of
collinear photons in QED or the emergence of a parton shower in QCD.
Thanks to the collinear enhancement, DM production at the LHC is typically
accompanied by the emission of several dark photons whose subsequent decay
leads to a collimated jet of SM particles.
These SM particles will often include electrons or muons, resulting in a
\emph{lepton jet} signature.  Depending on the dark photon decay length, the SM
particles in the lepton jets can be produced close to the primary interaction
vertex (prompt lepton jets) or further away from it (displaced
lepton jets).  In this paper, we explore both prompt and displaced lepton jets
from the theoretical and from the phenomenological side. In particular, we
derive semi-analytic expressions for the spectrum of radiated $\darkmed$
particles, and we compute limits on $U(1)'$ models from two ATLAS lepton jet
searches~\cite{Aad:2012qua, Aad:2014yea}. We also discuss how the discovery
potential will improve the 13~TeV LHC.

Lepton jets have been discussed previously in various contexts.  In
supersymmetric models, they can be a signal of cascade decays of heavy new
particles into much lighter states which subsequently decay to lepton
pairs~\cite{ArkaniHamed:2008qp, Cheung:2009su, Katz:2009qq, Bai:2009it,
Baumgart:2009tn, Chan:2011aa}. Lepton jets could also arise in models
featuring non-standard Higgs decays to dark sector particles that subsequently
decay to leptons~\cite{Han:2007ae, Falkowski:2010gv, Curtin:2013fra,Gupta:2015lfa}. DM
production in association with a single radiated $\darkmed$ boson have been discussed
in~\cite{Gupta:2015lfa, Autran:2015mfa}. If the $\darkmed$ is heavy, it can be
reconstructed as a dilepton resonance, while for light $\darkmed$, a narrow hadronic
jet from its decay to quarks can be a promising signature, especially since an
$\darkmed$-induced jet could be distinguished from a QCD jet by using jet
substructure techniques~\cite{Bai:2015nfa}.  In related works, multi-lepton
final states at $B$ factories have been studied in \cite{Batell:2009yf,
Essig:2009nc}.  Several recent papers discuss how strong dynamics in the dark
sector can lead to the formation of a dark parton shower, with subsequent decay
of dark sector particles into hadronic jets with non-standard properties like
displaced vertices, high multiplicities of vertices in the jet
cone~\cite{Schwaller:2015gea} and significant missing energy aligned with a
jet~\cite{Cohen:2015toa}. It is even possible that the DM itself
interacts relatively strongly with SM particles, so that its production at a
collider would lead to jets rather than the usual missing energy
signature~\cite{Bai:2011wy}. An experimental search for displaced hadronic jets
has been presented in \cite{Aad:2015uaa} by the ATLAS collaboration.

The plan of this paper is as follows: in sec.~\ref{sec:model}, we introduce our
simplified model of radiating dark matter and discuss its main phenomenological
features.  We then analyze the physics of dark photon showers in
sec.~\ref{sec:dark-shower} both analytically and numerically.  We derive the
expected spectrum of radiated dark photons in three different ways: using
recursive integral expressions, using an inverse Mellin transform to compute
the spectrum from its moments, and using a Monte Carlo simulation in
Pythia~8.2~\cite{Sjostrand:2014zea, Carloni:2010tw, Carloni:2011kk}.  Our main
phenomenological results are presented in sec.~\ref{sec:collider}. There, we
recast the existing ATLAS searches for prompt lepton jets~\cite{Aad:2012qua}
and for displaced lepton jets~\cite{Aad:2014yea} to set limits on our dark
radiation model, and we also compute the sensitivity of future LHC searches at
13~TeV center of mass energy.  We summarize our results and conclude in
sec.~\ref{sec:conclusions}.

\section{Model}
\label{sec:model}

\subsection{The Lagrangian}

The results of this paper can be applied to any model in which a
$\lesssim 10$~GeV DM particle interacts with a new $\lesssim 10$~GeV
gauge boson, as long as the typical energy at which DM
particles are produced at the LHC is much higher than their masses.
To illustrate our main points, we will use a
specific toy model of radiating DM, in which the dark sector consists of a fermionic DM
particle $\chi$ and a massive $U(1)'$ gauge boson $\darkmed$.  The
coupling strength between $\chi$ and $\darkmed$ is described by a dark
fine structure constant $\alpha_\darkmed \equiv g_\darkmed^2 / (4\pi)$.
Moreover, the dark photon $\darkmed$ is kinetically mixed with the SM
photon.  The Lagrangian for the dark sector is thus
\begin{align}
  \mathcal{L}_\text{dark} \equiv
      \bar{\chi} (i\slashed{\pd} - m_\chi + i g_\darkmed \slashed{\darkmed}) \chi
    - \frac{1}{4} F'_{\mu\nu} F'^{\mu\nu}
    - \frac{1}{2} m_\darkmed^2\darkmed_\mu\darkmed^\nu
    - \frac{\epsilon}{2} F'_{\mu\nu} F^{\mu\nu} \,,
  \label{eq:L-dark}
\end{align}
where $m_\darkmed$ and $m_\chi$ are the masses of $\darkmed$ and $\chi$,
respectively. We denote by $F^{\mu\nu}$ and $F'^{\mu\nu}$ the field strength
tensors of the SM photon and the dark photon $\darkmed$, respectively. The
dimensionless coupling constant $\epsilon$ describes the strength of the
kinetic mixing between the $\darkmed$ and the photon.
Due to
$SU(2)$ invariance, the $\darkmed$ should also mix kinetically with the $Z$,
but effects of this mixing are suppressed by factors of $m_\darkmed^2 / M_Z^2$
and are therefore neglected in the following.

Since the kinetic mixing between $\darkmed$ and the photon is too small to lead to
significant DM production at the LHC, we assume an additional DM--SM coupling.
For definiteness, we take this coupling to be through a heavy $s$-channel
vector resonance $\heavymed$ with couplings to all quark flavors and to dark
matter. Since the dynamics of dark radiation does not depend on the primary DM
production mechanism but only on the production cross section and to some
extent on the DM energy spectrum, our results will apply to any model in which
DM particles can be produced in significant numbers at the LHC, including for
instance models with contact interactions, $t$-channel mediators, or Higgs
portal interactions. The relevant terms in the Lagrangian of our toy model are
\begin{align}
  \mathcal{L}_\heavymed \equiv
      g_q \, \sum_f \bar{q}_f \slashed{\heavymed} q_f
    + g_\chi \, \bar{\chi} \slashed{\heavymed} \chi \,,
  \label{eq:L-Zprime}
\end{align}
where $q_f$ is the SM quark field of flavor $f$ and $g_q$, $g_\chi$ are the
$\heavymed$ couplings to quarks and DM particles, respectively. For simplicity,
we have assumed the coupling to quarks to be flavor universal.

\begin{figure*}
  \begin{center}
    \includegraphics[width=0.35\textwidth]{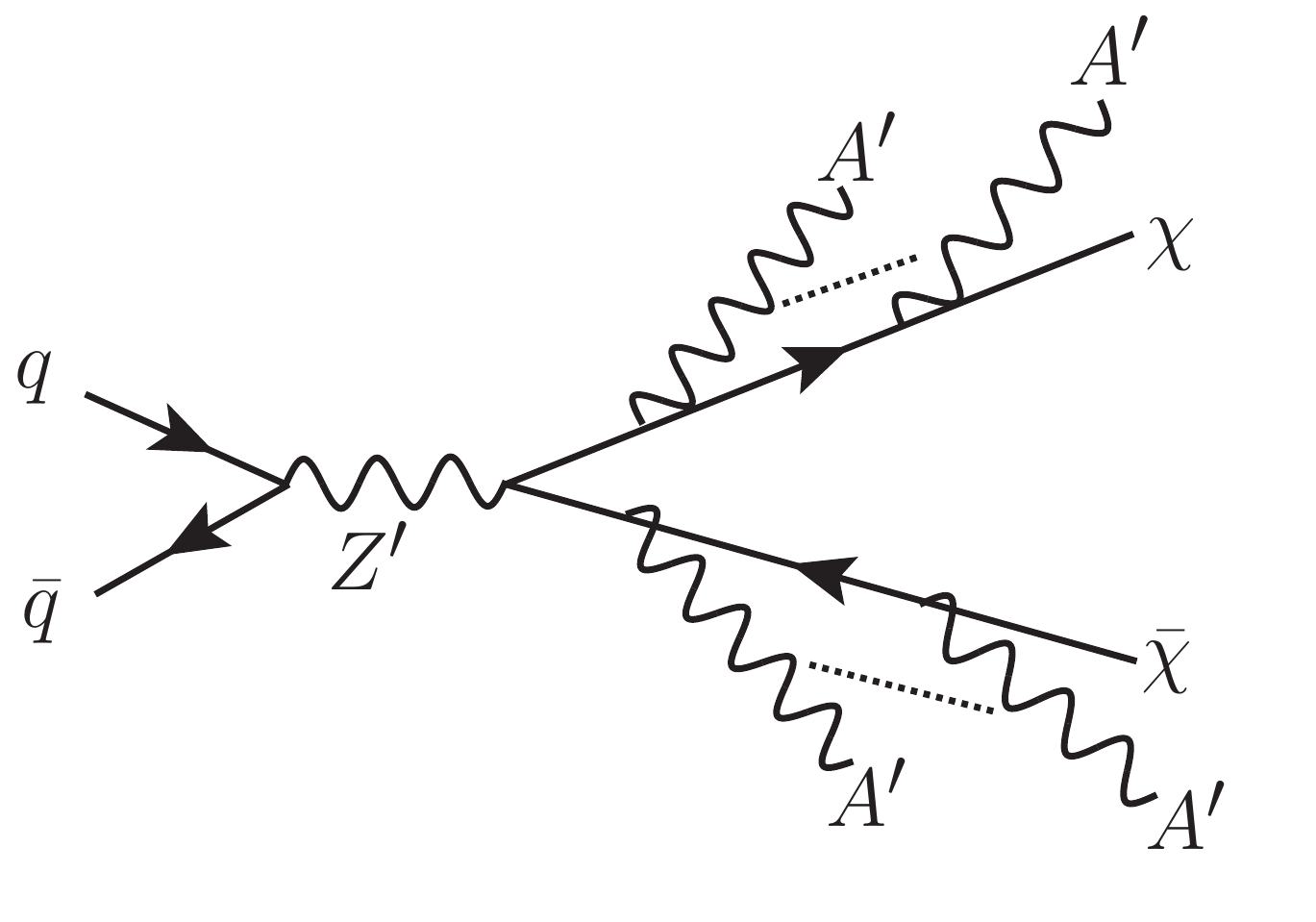}
  \end{center}
  \vspace{-0.5cm}
  \caption{Feynman diagram for dark matter pair production at the LHC through
    an $s$-channel $\heavymed$ resonance, followed by ``dark radiation'',
    i.e.\ emission of several--- mostly soft or collinear---dark photons.}
  \label{fig:SignalFD}
\end{figure*}

Fig.~\ref{fig:SignalFD} illustrates DM pair production at the LHC via
$s$-channel $\heavymed$ exchange, followed by radiation of several dark photons.
Due to the assumed lightness of $\chi$ and $\darkmed$, the emission
is enhanced in the collinear direction. Due to this enhancement, with a
moderate dark fine structure constant $\alpha_\darkmed \sim \mathcal{O}(0.1)$, we typically
expect a few dark photons to be radiated in each DM pair production process. We
will discuss the dynamics of this ``dark radiation'' in
sec.~\ref{sec:dark-shower} using a formalism analogous to parton showers in
QCD.

\subsection{Benchmark points}

We define two benchmark points in the parameter space of our
toy model, which we will use to illustrate our main points in
secs.~\ref{sec:dark-shower} and \ref{sec:collider}.  In
sec.~\ref{sec:collider}, we will also discuss in detail how departing from
these benchmark points affects our results.  The two benchmark points
A and B are summarized in table~\ref{tab:Benchmark}, together with
several phenomenological observables derived from them.

\begin{table}
  \begin{ruledtabular}
  \begin{tabular}{c|ccccccc|cccccccc}
      & $m_\heavymed$ & $g_q$ & $g_\chi$ & $m_\chi$ & $m_\darkmed$ & $\alpha_\darkmed$ & $c\tau$ & $\epsilon$ &
             $\sigma_7${\footnotesize($\heavymed$)} & $\sigma_8${\footnotesize($\heavymed$)} & $\sigma_{13}${\footnotesize($\heavymed$)} &
             $\Gamma_\heavymed$ & BR{\footnotesize($\heavymed\!\to\!\chi\bar\chi$)} & $2\!\ev{n_\darkmed}_{\!8}$ & $2\!\ev{n_\darkmed}_{\!13}$\\
      & [TeV]       &     &        & [GeV]  & [GeV]      &     & [mm] & [$10^{-6}$] & [pb]  & [pb]  & [pb]  & [GeV] &         &      &      \\
    \hline
    A & 1           & 0.1  & 1     & 4      & 1.5        & 0.2 & 10   & 2.8         & 0.58  & 0.85  & 2.7   & 31.3  & 84.8\%  & 3.50 & 3.51 \\
    B & 1           & 0.03 & 0.3   & 0.4    & 0.4        & 0.2 & 1    & 24          & 0.052 & 0.076 & 0.244 & 2.82  & 84.8\%  & 5.15 & 5.17 \\
  \end{tabular}
  \end{ruledtabular}
  \caption{
    Values of the model parameters $m_\heavymed$ (heavy mediator mass), $g_q$,
    $g_\chi$ (heavy mediator couplings), $m_\chi$ (DM mass), $m_\darkmed$ (dark
    photon mass), $\alpha_\darkmed$ (dark fine structure constant) and $c\tau$
    (dark photon decay length) at our two benchmark points A and B.  We also
    show the resulting values for several derived quantities, in particular the
    kinetic mixing $\epsilon$ corresponding to the given $c\tau$ and $m_\darkmed$,
    the resonance cross sections $\sigma_7(\heavymed)$, $\sigma_8(\heavymed)$
    and $\sigma_{13}(\heavymed)$ for $\heavymed$ production $p p \to \heavymed$
    at the 7~TeV, 8~TeV and 13~TeV LHC, respectively, the total decay width
    $\Gamma_\heavymed$ of the heavy mediator, its branching ratio to DM pairs,
    and the average numbers $2\ev{n_\darkmed}_8$ ($2\ev{n_\darkmed}_{13}$) of
    radiated dark photons in each DM pair production event at the 8~TeV
    (13~TeV) LHC.}
  \label{tab:Benchmark}
\end{table}

In both cases, we assume a $\heavymed$ mass of $1$~TeV.  We choose $g_q$ and
$g_\chi$ such that the resonant $\heavymed$ production cross section is about
1~pb at the 8~TeV LHC for benchmark point~A, and about a factor of 10 smaller
for benchmark point~B. In both cases, the $\heavymed$ has a branching ratio
$\sim 85\%$ for the decay $\heavymed \to \bar\chi \chi$.  We choose both
$m_\chi$ and $m_\darkmed$ to be of order GeV or below. Together with a
moderately large dark fine structure constant $\alpha_{\darkmed} = 0.2$, this
leads to the radiation of a significant number of $\darkmed$ bosons when DM is
produced at the LHC.  Note that the number of radiated $\darkmed$ bosons
is almost independent of the collider center of mass energy because the boost
of the primary DM particles, which determines the amount of radiation,
is mostly dictated by the $s$-channel $\heavymed$ resonance.
We choose the dark photon decay length $c\tau$ such that
at benchmark point A, most dark photons will decay far from their production
vertex, while at benchmark point B, the displacement of the decay vertex is
much smaller.  Thus, benchmark point~A is in the realm of lepton jet searches
requiring displaced decay vertices, while benchmark point~B is most easily
detectable in searches for prompt lepton jets. Note that in
table~\ref{tab:Benchmark} we treat $c\tau$ as a fundamental parameter and
$\epsilon$ as a derived quantity.  This will only be relevant when we vary the
model parameters around their benchmark values in sec.~\ref{sec:collider}
(figs.~\ref{fig:CLs1} and \ref{fig:CLs2}).  To make the discussion there more
transparent, we choose to keep $c\tau$ rather than $\epsilon$ fixed when
varying the other model parameters.  Of course, as long as all other model
parameters are fixed, there is a one-to-one correspondence between $c\tau$ and
$\epsilon$.

Our parameter choices are consistent with existing collider constraints: In
particular, dijet searches do not exclude $\heavymed$ bosons with the
parameters chosen here~\cite{Aad:2014aqa, Khachatryan:2015sja}.  For instance,
the ATLAS dijet search in $20.3$~fb$^{-1}$ of 8~TeV LHC data~\cite{Aad:2014aqa}
restricts the product $\sigma(\heavymed) \times \BR(\heavymed \to qq) \times A$
of the $Z'$ production cross section, the branching ratio to a quark--antiquark
final state, and the experimental acceptance factor $A \sim 0.6$ to be
$\lesssim 1$~pb at 95\% confidence level (CL). The corresponding CMS search
\cite{Khachatryan:2015sja} does not explicitly quote a limit at $m_\heavymed =
1$~TeV, but noting that no spectral feature is observed at this mass, the
limits given on heavier $\heavymed$ bosons can be extrapolated down, indicating
that the CMS limit is stronger than the ATLAS limits by perhaps a factor of
two.  The acceptance factor is similar in ATLAS and CMS.  Limits on dijet
resonances are also provided by the CDF experiment at the Tevatron, yielding
$\sigma(\heavymed) \times \BR(\heavymed \to qq) \times A \lesssim
0.11$~pb~\cite{Aaltonen:2008dn}.  Since CDF requires both jets to have a
rapidity $|y| < 1$, we conservatively estimate the acceptance in CDF is not
larger than in CMS.  Including an acceptance factor of 0.6, we predict at our
benchmark point~A $\sigma(\heavymed) \times \BR(\heavymed \to qq) \times A =
0.08$~pb at the 8~TeV LHC and $\sigma(\heavymed) \times \BR(\heavymed \to qq)
\times A = 0.001$~pb at the Tevatron, which is clearly consistent with the
above constraints. The $pp \to \heavymed \to \bar\chi\chi$ production cross
sections at benchmark point~B are even smaller.

DM production through heavy resonances is also constrained by monojet
searches~\cite{Aad:2015zva, Khachatryan:2014rra}.  ATLAS sets a lower limit
$m_\heavymed / \sqrt{g_q g_\chi} \gtrsim 2$~TeV~\cite{Aad:2015zva},
whereas we have $m_\heavymed / \sqrt{g_q g_\chi} \sim 3.2$~TeV at
benchmark point~A and $m_\heavymed / \sqrt{g_q g_\chi} \sim 10$~TeV
at benchmark point~B. Both benchmark points thus satisfy the limit.

We also need to consider cosmological constraints on the dark sector of our toy
model. The most important question here is whether $\chi$ can be a thermal
relic and account for all of the DM in the Universe. If there is no
particle--antiparticle asymmetry in the dark sector, the answer is no. The
reason is that at our benchmark points the annihilation cross
section~\cite{Liu:2014cma}
\begin{align}
  \ev{\sigma v}_{\bar\chi \chi \to \darkmed\darkmed}
    \simeq \frac{\pi \alpha_\darkmed^2}{m_\chi^2}
    \frac{(1 - m_\darkmed^2 / m_\chi^2)^{3/2}}{(1 - \frac{1}{2} m_{\darkmed}^2 / m_\chi^2)^2}
\end{align}
is much larger than the thermal relic cross section of $\text{few} \times 10^{-26}\
\text{cm}^3/\text{sec}$.  Reducing the annihilation cross section would either
require a significantly smaller $\alpha_\darkmed$, precluding significant dark
radiation at the LHC, or a larger $\darkmed$ mass, in particular $m_\darkmed >
m_\chi$. In the latter case, only the annihilation channel $\bar\chi \chi \to
\bar{q} q$ would remain open. The cross section for this channel
is~\cite{Zheng:2010js, Dreiner:2012xm}
\begin{align}
  \ev{\sigma v}_{\bar\chi\chi \to \bar{q} q}
    \simeq \frac{3 N_f g_q^2 g_\chi^2 }{2\pi}
           \frac{2 m_\chi^2 + m_q^2}{(4 m_\chi ^2 - m_{Z'}^2)^2}
           \sqrt{1 - \frac{m_q^2}{m_\chi^2}} \,,
\end{align}
where $N_f$ is the number of kinematically accessible quark flavors.
Since $\ev{\sigma v}_{\bar\chi\chi \to \bar{q} q}$ scales as
$g_q^2 g_\chi^2 m_\chi^2 / m_\heavymed^4$, it is much smaller than
the thermal relic value for the small $\chi$ masses that we are interested in
here. It would thus lead to overclosure of the Universe.
For symmetric DM models, we therefore conclude that $\chi$ must be a
subdominant component of the DM in the Universe~\footnote{There is an exception when $\darkmed$ mass is slightly higher than $\chi$ mass, one can exponentially reduce the thermal averaged annihilation cross section and achieve the right relic abundance \cite{D'Agnolo:2015koa}.}. This, of course, does not
change the LHC phenomenology that we are interested in here, it only means that
our toy model cannot provide a complete description of the dark sector.  The
fact that for symmetric DM models a discovery of $\chi$ through a dark
radiation signature is only possible if $\chi$ is a subdominant component of DM
is in line with the general statement that, in models with multiple thermally
produced DM components, it is likely that the subdominant DM components are
discovered at the LHC first due to their larger interaction strengths.  For
asymmetric DM, i.e.\ DM with a primordial particle--antiparticle asymmetry
generated at some high scale, $\chi$ could account for all of the DM in
the Universe. In this case, the large annihilation cross section to $\darkmed$
pairs is not a problem but a feature because it allows for efficient
annihilation of the symmetric component of the primordial DM soup. However another opposite charge particle might be needed due to gauge invariance for such light $\darkmed$ \cite{Petraki:2014uza}. Let us emphasize that, in this paper, we will indiscriminately call $\chi$ the dark matter, even though the above discussion shows that it does not need to be the dominant DM component.

Next, let us comment on constraints from direct and indirect DM searches.
Without a primordial $\chi$--$\bar\chi$ asymmetry, both types of constraints
are easily satisfied due to the required small relic abundance of $\chi$.  For
asymmetric DM, indirect searches are also insensitive because $\chi \bar\chi$
annihilation is impossible in such models. The sensitivity of direct detection
experiments is limited due to the small $m_\chi$ at our benchmark points. For
benchmark point B with $m_\chi = 0.4$~GeV, the DM mass is obviously below the
direct detection threshold, while for benchmark point A with $m_\chi = 4$~GeV,
one might worry about limits from low-threshold experiments.  The best limit in
this mass range comes from CDMSlite and requires the spin-independent
$\chi$--nucleon scattering cross section $\sigma_{\chi N}$ to be below $1.5
\times 10^{-40}$~cm$^2$ if $\chi$ accounts for all of the DM in the
Universe~\cite{Agnese:2013jaa}. Our prediction at benchmark point A is
$\sigma_{\chi N} \simeq 6.5 \times 10^{-42}$~cm$^2$, avoiding the bound by more
than an order of magnitude.

A different set of bounds on our model comes from the fact that $\chi$
particles have self-interactions, mediated by the $\darkmed$ boson.
Such DM self-interactions are constrained by observations of colliding galaxy
cluster (most notably the Bullet Cluster)~\cite{Randall:2007ph} and
by measurements of the ellipticity of DM halos of groups of
galaxies~\cite{Peter:2012jh}. The resulting limit on the DM self-scattering
cross-section is
\begin{align}
  \sigma_{\chi\chi} / m_\chi  \lesssim 1\ \text{cm}^2 /\text{g}
                              = 1.78 \times 10^{-24}\ \text{cm}^2 / \text{GeV} \,.
\end{align}
Obviously, this constraint applies only if $\chi$ forms the bulk of the
DM in  the Universe. As explained above, this is only possible in our scenario
if there is a particle--antiparticle asymmetry in the dark sector.
In the perturbative regime, where $\alpha_{A'} m_\chi  / m_{A'} \lesssim 1$,
a simple calculation shows that the predicted non-relativistic
self-scattering cross section is~\cite{Tulin:2013teo}
\begin{align}
  \sigma_{\chi\chi} / m_\chi \simeq
  5 \times 10^{-31}\ \text{cm}^2/\text{GeV} \times
    \bigg( \frac{\alpha_{A'}}{0.1} \bigg)^2
    \bigg( \frac{m_\chi}{\text{GeV}} \bigg)
    \bigg( \frac{\text{GeV}}{m_{A'}} \bigg)^4 \,.
\end{align}
For our benchmark point A (B) from table~\ref{tab:Benchmark}, this leads to
$\sigma_{\chi\chi}/m_\chi \simeq 10^{-30}\ \text{cm}^2/\text{GeV}$
$(10^{-29}\ \text{cm}^2/\text{GeV})$,
well within the observational limit.

Regarding the dark photon mass $m_\darkmed$ and its kinetic mixing $\epsilon$
with the SM photon, strong constraints exist from low energy experiments and
from astrophysics. These constraints require $\epsilon \lesssim 10^{-10}$ for
$m_\darkmed \lesssim 10$~MeV, but relax to $\epsilon \lesssim 10^{-3}$ for $10\
\text{MeV} \lesssim m_\darkmed \lesssim 10\ \ \text{GeV}$, the mass range that
we are interested in here.  Our benchmark points from table~\ref{tab:Benchmark}
are thus consistent with limits from dark photon searches. We will discuss
these limits in more detail in sec.~\ref{sec:collider}, see in particular
fig.~\ref{fig:CLs2D}.

\subsection{Dark Photon decay}

\begin{figure*}
  \begin{center}
    \includegraphics[width=\textwidth]{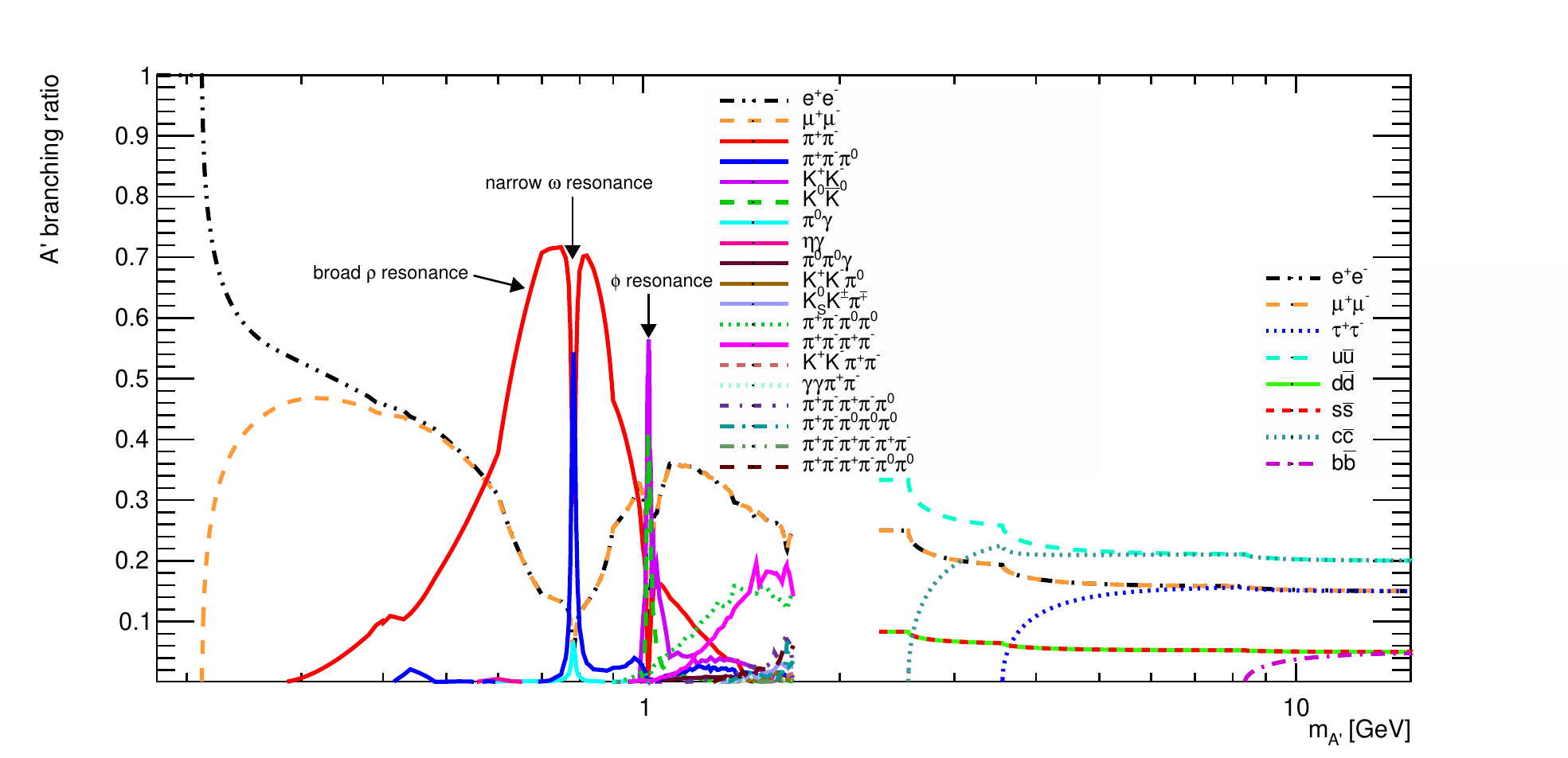}
  \end{center}
  \vspace{-0.5cm}
  \caption{Branching ratios for the 19 dark photon decay channels included in our
    analysis as a function of the dark photon mass.}
  \label{fig:BR}
\end{figure*}

After dark photons have been radiated from pair-produced $\chi$ particles
at the LHC, they decay to SM particles via their kinetic mixing
with the photon (see eq.~\eqref{eq:L-dark}). For
$m_\darkmed \lesssim 2~\text{GeV}$ and $m_\darkmed < 2 m_\chi$, the
dominant decay modes include $\darkmed \to e^+ e^-, \; \mu^+\mu^-$, as
well as decays to various hadronic resonances.  The partial decay
width to leptons of a specific flavor $\ell$ is
\begin{align}
  \Gamma(\darkmed \to \ell^+\ell^-) =
    \frac{1}{3} \alpha \epsilon^2 m_\darkmed
    \sqrt{1 - 4 \frac{m_\ell^2}{m_\darkmed^2}}
    \bigg(1 + 2 \frac{m_\ell^2}{m_\darkmed^2} \bigg) \,.
  \label{eq:DecayWidth}
\end{align}
The hadronic decay width can be calculated using measurements of hadron
production cross sections at $e^+ e^-$ colliders, in particular the ratio $R(s)
\equiv \sigma(e^+e^- \to \text{hadrons}) / \sigma(e^+e^- \to \mu^+\mu^-)$,
where $s$ is the center-of-mass energy.  At $s = m_\darkmed^2$, the ratio
between the partial $\darkmed$ decay width into a given hadronic final state
and the partial decay with to the $\mu^+\mu^-$ final state should be equal to
$R(s)$:
\begin{align}
  \Gamma(\darkmed \to \text{hadrons}) =
    \Gamma(\darkmed \to \mu^+\mu^-) \, R(s=m_\darkmed^2) \,.
\end{align}
We use hadronic cross-sections from refs.~\cite{Whally:2003, hepdata} to
calculate the $\darkmed$ decay branching ratios.  We include 19 decay channels,
shown in fig.~\ref{fig:BR}. In the calculation, care must be taken to avoid
double counting of hadronic degrees of freedom.  For example, the decay
$\darkmed \to \pi^+\pi^- \pi^0 \pi^0$ receives contributions both from the
direct production of pions and from the decay $\darkmed \to \omega \pi^0$,
followed by $\omega \to \pi^+ \pi^- \pi^0$. We choose to preferentially include
channels with pions and kaons. For heavier mesons, we only include decay
channels in which there is no double counting.  In principle, there is also a
double counting problem between kaon and pion final states because of the fully
hadronic kaon decay channels.  Fortunately, these channels have very small
branching ratios and can be neglected.

For $m_\darkmed \gtrsim 2$~GeV, hadronic $\darkmed$ decays are more
conveniently described in QCD language as decays to quark pairs.  The partial
decay width to a quark pair of a given flavor $q_f$ is
\begin{align}
  \Gamma(\darkmed \to q_f\bar{q_f}) =
    N_c \, Q_{q_f}^2 \, \Gamma(\darkmed \to \ell^+\ell^-)|_{m_\ell=m_{q_f}} \,,
\end{align}
where $Q_{q_f}$ is the electric charge of $q_f$ and $N_c = 3$ is a color
factor.  At intermediate $\darkmed$ masses $\sim 2$~GeV, a very large number of
hadronic resonances appears, so that neither the QCD description in terms of
quark final states nor the low energy effective theory involving just a few
hadronic states are applicable. We therefore do not consider this mass range in
the following, but we will see in sec.~\ref{sec:collider} that our results on
the LHC sensitivity can be smoothly interpolated between the QCD regime and the
low-energy regime.

\section{Dark Parton Shower}
\label{sec:dark-shower}

\begin{figure}
  \centering
  \includegraphics[width=0.3\textwidth]{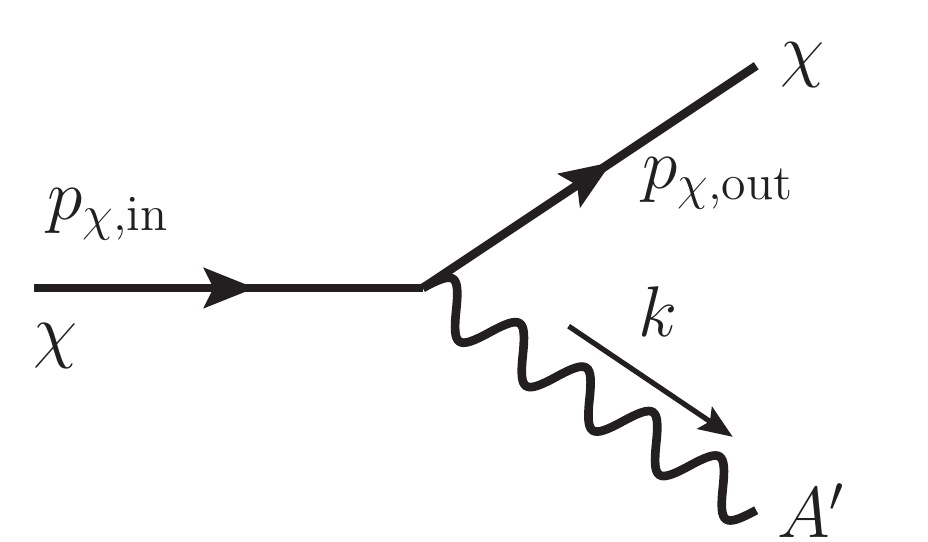}
  \caption{Radiation of a single dark photon $\darkmed$ from a DM particle $\chi$.}
  \label{fig:splitting}
\end{figure}

Before presenting our numerical results, let us provide some analytic
discussion of dark photon radiation. Part of the following discussion is
based on refs.~\cite{Plehn:2009nd, Ciafaloni:2010ti}.  We first note that
factorization theorems ensure that the cross section can be factorized into
short range (hard process) and long range (shower)
contributions~\cite{Collins:1989gx}.  Thus, we can treat the dark photon shower
as independent of the primary $\bar\chi\chi$ production process.  The relevant
processes in the development of an $\darkmed$ shower are $\chi \to \chi +
\darkmed$ and $\darkmed \to \bar\chi \chi$.  In the following, we include only
the first process, which is shown also in fig.~\ref{fig:splitting}.  Its
probability is much larger than the probability for $\darkmed \to \bar\chi \chi$
due to the structure of the respective splitting kernels. In particular, the splitting
kernel for $\darkmed \to \bar\chi\chi$ is not divergent in the soft limit, in
analogy to photon splitting in QED.  The radiation of gauge bosons from
fermions, on the other hand, is formally divergent in the soft and collinear
phase space regions.  In our scenario, the dark matter and dark photon masses
regularize the radiation probability in these regions by providing a natural
infrared cut-off. From a phenomenological point of view, we are more interested
in collinear emission than in soft emission because the decay products of very
soft dark photons would not be detectable at the LHC.  We work under the
assumption of strongly ordered emission, i.e.\ the virtuality of the incoming
particle in a splitting process is assumed to be much larger than the
virtuality of the outgoing particles. The latter are therefore treated as on-shell.
Within the strongly ordered emission assumption, we can consider each splitting
$\chi \to \chi + \darkmed$ as an isolated process, since a secondary splitting
would be a small perturbation with respect to the first one.

The differential probability for a collinear splitting can be written as
\begin{align}
  \frac{\alpha_{\darkmed}}{2\pi} dx \, \frac{dt}{t} P_{\chi\to\chi}(x, t) \,,
  \label{eq:splitting-probability}
\end{align}
where $t$ is the virtuality of the incoming DM particle and $x$ is the fraction of its
energy that is transferred to the outgoing DM particle.  The
quantity $P_{\chi\to\chi}(x, t)$ is the splitting kernel that encodes
the model-dependent details of the splitting probability.
We call attention to the fact that we treat $\alpha_\darkmed$ as independent of $t$,
i.e.\ we neglect renormalization group running.
Since the running is logarithmic, it will not play a major role in the development
of the dark photon shower.  Moreover, the running would depend on the full particle content
of the dark sector and is thus highly model dependent.
Eq.~\eqref{eq:splitting-probability} can be
understood as the squared matrix element of the splitting process,
multiplied by the propagator of the incoming DM particle and
integrated over the phase space of the outgoing particles.  In the
limit $t \to 0$ (highly collinear splitting), the propagator factor
$1/t$ in eq.~\eqref{eq:splitting-probability} diverges as expected. It is
prevented from going all the way to zero thanks to the
non-zero dark photon mass.

\subsection{Kinematics and notation}
\label{sec:ds-kinematics}

We write the off-shell 4-momentum of the incoming DM
particle in a given splitting process as (see ref.~\cite{Ciafaloni:2010ti} for
a slightly different approach)
\begin{align}
  p_{\chi, \text{in}} = (E, 0, 0, p) \,.
\end{align}
Notice that in general $E \neq \sqrt{\hat s}/2$. For instance, if the second
$\chi$ particle, which we call the spectator, is on-shell and does not radiate,
energy momentum conservation requires its three-momentum to be $(0,0,-p)$.
This fixes the energy of the spectator to be $E_s^2 = p^2 + m_\chi^2$, and
hence $E \geq E_s + m_\darkmed$ if at least one dark photon is radiated in the
event. In the following, we will nevertheless assume $E = E_s = \sqrt{\hat
s}/2$, the so called collinear approximation.  Naturally, we expect it to fail
when $m_\darkmed + m_{\chi} \sim E$, or when the opening angle between the
splitting products is large, so the splitting is not collinear any more.

The on-shell 4-momenta of the outgoing DM particle and dark photon after the
splitting are
\begin{align}
  p_{\chi,\text{out}} &= \big( x E,\; -k_t,\; 0,\;
                           \sqrt{x^2 E^2 - k_t^2 - m_\chi^2} \big),  \\
\intertext{and}
  k &= \big( (1-x) E,\; k_t,\; 0,\; \sqrt{(1-x)^2 E^2 - k_t^2 - m_\darkmed^2} \big) \,,
\end{align}
respectively.
Since the energies of the outgoing particles must be larger than their
masses and because energies and momenta
have to be positive, the following constraints must be fulfilled:
\begin{gather}
  m_\chi/E < x < 1-m_\darkmed/E \,,
  \qquad
  x^2 E^2 - m_\chi^2 \ge k_t^2 \,,
  \qquad
  (1-x)^2 E^2 - m_\darkmed^2 \ge k_t^2
  \,.
  \label{eq:constraints}
\end{gather}
The first relation defines in particular the allowed range of the variable
$x$ in the splitting probability eq.~\eqref{eq:splitting-probability}.
Unfortunately, this range depends on $E$ and thus on all preceding
splitting processes. This would preclude us from treating the dark photon
shower analytically. Therefore, we will in the following take the
minimum and maximum values of $x$ to be
\begin{align}
  x_{\min} \equiv m_\chi / E_0 \,,
  \qquad
  x_{\max} \equiv 1 - m_\darkmed / E_0 \,,
  \label{eq:xmin-xmax}
\end{align}
where $E_0$ is the initial energy of the DM particle $\chi$ before emission of
the first dark photon. Since most emitted dark photons are soft, we expect $E$
to be close to $E_0$ for most splittings.  The relations \eqref{eq:xmin-xmax}
would even be exact for $m_\chi = m_\darkmed = 0$, but since we have $m_\chi$,
$m_\darkmed \ll E_0$, we expect them to be very good approximations also in our
case.

The virtuality $t$ is defined as
\begin{align}
  t \equiv (p_{\chi,\text{out}}+k)^2 - m_\chi^2
    =      m_\darkmed^2 + 2 p_{\chi,\text{out}} \cdot k\,,
\end{align}
and the splitting kernel is given by~\cite{Catani:2002hc}
\begin{align}
  P_{\chi\to\chi}(x, t) = \frac{1+x^2}{1-x} - \frac{2(m_\chi^2+m_\darkmed^2)}{t} \,.
  \label{eq:massivekernel}
\end{align}
We see that $P_{\chi\to\chi}(x, t)$ becomes large for $x \to 1$, where the
emitted dark photon is very soft. Since we are dealing with massive dark
photons, $x$ is kinematically required to always be $< 1$ (see eqs.~\eqref{eq:constraints}
and \eqref{eq:xmin-xmax}). In the following, we will neglect the
mass dependent term in $P_{\chi\to\chi}$ and call the thus approximated splitting
kernel $P_{\chi\to\chi}(x)$.  Like the approximation used for $x_{\min}$ and
$x_{\max}$ above, also this approximation is necessary to make different
splittings completely independent of each other, a precondition for our
analytic calculations.

In order to also regularize the collinear
$t \to 0$ divergence in eq.~\eqref{eq:splitting-probability},
care must be taken also when choosing the integration limits for $t$.
The lower bound $t_{\min}$ for $t$ is obtained when $k_t \to 0$, that is
\begin{align}
  t_{\min}(x)
      &= m_\darkmed^2 + 2 p_{\chi,\text{out}} \cdot k \big|_{k_t \to 0} \nonumber\\
      &= m_\darkmed^2 + 2 \Big(
             E^2 x (1-x) - \sqrt{x^2 E^2 - m_\chi^2}
             \sqrt{(1-x)^2 E^2 - m_{A^\prime}^2} \Big) \,,
  \label{eq:tmin}
\end{align}
while the upper bound is obtained when $k_t$ is maximal, within the
constraints~\eqref{eq:constraints}, that is,
\begin{align}
  t_{\max}(x) &=
    m_{A^\prime}^2 + 2 p_{\chi,\text{out}} \cdot k \big|_{k_{t,\text{max}}} \,,
  \label{eq:tmax}
\intertext{with}
  k_{t, \text{max}}^2 (x) &= \min \big\{ (1 - x)^2 E^2 - m_{A^\prime}^2 \,,\
                                     x^2 E^2 - m_\chi^2 \big\} \,.
\end{align}
Interestingly, when $x$ takes its maximum or minimum value allowed by
eq.~\eqref{eq:constraints}, $k_{t, {\max}}$ becomes $0$, which leads $t_{\min}
= t_{\max}$ and the splitting probability goes to $0$.  As in
eq.~\eqref{eq:xmin-xmax}, we will in the following take $E \to E_0$ also when
evaluating $t_{\min}$ and $t_{\max}$. This approximation is necessary to make
consecutive splitting processes completely independent and thus treat the dark
photon shower analytically.

\subsection{Number of emitted dark photons}
\label{sec:ds-n}

A DM particle is produced at the LHC with a certain maximal virtuality
$t_{\max}$ and then radiates in general several dark photons. In each
splitting, its virtuality is reduced, until it finally reaches the
infrared cutoff $t_{\min}$. The expectation value for the
number of radiated dark photons is
\begin{align}
  \ev{n_\darkmed} \simeq \frac{\alpha_\darkmed}{2\pi}
           \int_{x_{\min}}^{x_{\max}} \! dx \, \int_{t_{\min}}^{t_{\max}} \frac{dt}{t}
           P_{\chi\to\chi}(x) \,.
  \label{eq:ev-n}
\end{align}
Note that this compact expression is valid only if we neglect the
$t$-dependence of the splitting kernel and of the integration boundaries
$x_{\min}$, $x_{\max}$, $t_{\min}$, $t_{\max}$ defined in the previous section.

The probability that a DM particle $\chi$ radiates exactly $m$ dark photons
is given by a Poisson distribution~\cite{Plehn:2009nd}:
\begin{align}
  p_m = \frac{e^{-\ev{n_\darkmed}} \ev{n_\darkmed}^m}{m!} \,.
  \label{eq:pm}
\end{align}
To see this, note that eq.~\eqref{eq:splitting-probability} implies that the
probability for \emph{no} splitting to occur between $t_{\max}$ and $t_{\min}$
is given by the Sudakov factor
\begin{align}
  \Delta(t_{\min}, t_{\max}) \equiv e^{-\ev{n_\darkmed}} \,,
  \label{eq:Sudakov}
\end{align}
with $\ev{n_\darkmed}$ defined in eq.~\eqref{eq:ev-n}. The probability for
exactly one splitting is then
\begin{align}
  p_1
    &= \frac{\alpha_\darkmed}{2\pi}
       \int_{x_{\min}}^{x_{\max}}\! dx \, \int_{t_{\min}}^{t_{\max}} \frac{dt}{t}
       \Delta(t_{\min}, t) P_{\chi\to\chi}(x) \Delta(t, t_{\max}) \nonumber\\
    &= e^{-\ev{n_\darkmed}} \ev{n_\darkmed} \,.
\end{align}
In the first line, the factors $\Delta(t_{\min}, t)$ and
$\Delta(t, t_{\max})$ give the probabilities for no splittings to happen
in the intervals $[t_{\min}, t)$ and $(t, t_{\max}]$, respectively.
The splitting kernel $P_{\chi\to\chi}(x)$ describes the probability that
a splitting happens at virtuality $t$.
For two splittings, we have in analogy
\begin{align}
  p_2
    &= \bigg( \frac{\alpha_\darkmed}{2\pi} \bigg)^2
       \int_{x_{\min}}^{x_{\max}}\! dx \,
       \int_{t_{\min}}^{t_{\max}} \frac{dt}{t}
       \int_{x_{\min}}^{x_{\max}}\! dx' \,
       \int_{t_{\min}}^t \frac{dt'}{t'}
       \Delta(t_{\min}, t) P_{\chi\to\chi}(x) \Delta(t, t') P_{\chi\to\chi}(x')
       \Delta(t', t_{\max}) \nonumber\\
    &\simeq e^{-\ev{n_\darkmed}} \frac{1}{2!} \bigg( \frac{\alpha_\darkmed}{2\pi} \bigg)^2
       \int_{x_{\min}}^{x_{\max}}\! dx \,
       \int_{t_{\min}}^{t_{\max}} \frac{dt}{t}
       \int_{x_{\min}}^{x_{\max}}\! dx' \,
       \int_{t_{\min}}^{t_{\max}} \frac{dt'}{t'}
       P_{\chi\to\chi}(x) P_{\chi\to\chi}(x') \nonumber\\
    &= e^{-\ev{n_\darkmed}} \frac{\ev{n_\darkmed}^2}{2!} \,.
  \label{eq:p2}
\end{align}
In the second line, we have extended the integration region of the
$t'$ integral to the full range $[t_{\min}, t_{\max}]$ and included a
factor $1/2!$ to compensate for the double counting induced that way.
This is the place where the approximations discussed in
sec.~\ref{sec:ds-kinematics} are crucial: extending the integration interval
is only possible if we can assume that $x_{\min}$, $x_{\max}$, $t_{\min}$ and
$t_{\max}$ are the same for all splittings and that $P_{\chi\to\chi}$ is
independent of $t$. Eq.~\eqref{eq:p2} straightforwardly generalizes to
larger numbers of splittings, yielding the Poisson probabilities $p_m$
in eq.~\eqref{eq:pm}.

\subsection{Recursion relation for the $\chi$ and $\darkmed$ energy spectra}
\label{sec:ds-recursion}

Let us now study the energy distributions of $\chi$ and $\darkmed$ in events
with radiation of multiple $\darkmed$ bosons.  Our starting point is the
observation that, in the collinear limit, each splitting process is
independent.  As explained in sec.~\ref{sec:ds-kinematics}, we will make the
crucial approximation that the integration boundaries $x_{\min}$, $x_{\max}$,
$t_{\min}$, $t_{\max}$ are the same for each emission process and are given by
eqs.~\eqref{eq:xmin-xmax}, \eqref{eq:tmin} and \eqref{eq:tmax} (with $E \to
E_0$). We moreover neglect again the $t$-dependence of the splitting kernel
$P_{\chi\to\chi}$.  These approximations will allow us to compute the energy
distributions recursively.

If we call the final energy of the on-shell DM particle after final state
radiation $E_\chi$ and the energy
fraction retained by the DM particle after all splittings $X \equiv E_\chi /
E_0$,\footnote{We use capital letters to denote quantities normalized to the
  initial energy $E_0$ of the DM particle before the first splitting, and lower case
  letters for quantities normalized to the energy of the DM particle as it
  enters a particular splitting process during the development of the
$\darkmed$ shower.} we can write the final DM energy distribution as
\begin{align}
  f_\chi(X) = \sum_{m=0}^\infty p_m \, f_{\chi, m}(X)
  \label{eq:chiPDF} \,,
\end{align}
where $f_{\chi, m}(X)$ is the energy distribution of $\chi$ in events with
exactly $m$ emitted dark photons.  $f_{\chi, m}(X)$ obeys the recursion
relation (see also~\cite{Arbuzov:1999cq})
\begin{align}
  f_{\chi, m+1}(X) = \int_{x_{\min}}^{x_{\max}} \!
    dx_m \, f_{\chi, 1}(x_m) \, \frac{f_{\chi, m}(X/x_m)}{x_m} \,
    \Theta(x_{\min} \le X \le x_{\max}) \,,
  \label{eq:recursion-chi}
\end{align}
where
\begin{align}
  f_{\chi, 1}(X) \equiv \frac{1}{\ev{n_\darkmed}}
           \frac{\alpha_{A^\prime}}{2\pi} \,
           \int_{t_{\min}}^{t_{\max}} \frac{dt}{t}
           P_{\chi\to\chi}(X) \; \Theta(x_{\min} \le X \le x_{\max}) \,,
  \label{eq:fchi-1}
\end{align}
is the energy fraction for dark radiation showers with only a single splitting
and $\Theta(\,\cdot\,)$ denotes a window function that is 1 when the condition in
parentheses is satisfied and 0 otherwise. Note that for $m=1$, the definitions of
$x$ as the fraction of energy the DM particle retains in a single splitting and
of $X$ as the fraction of energy the DM particle retains after \emph{all} splittings
are identical. For $m=0$, we define $f_{\chi,
0}(X) \equiv \delta(1-X)$.  Note that $f_{\chi, m}(X)$ is normalized according
to the condition
\begin{align}
  \int \! dX \, f_{\chi, m}(X) = 1 \,.
  \label{eq:fchi-m-norm}
\end{align}
When evaluating $f_\chi(X)$ numerically, we truncate the recursive
series at $m = 10$, which is justified by the fact that the Poisson probability
$p_m$ is very small at $m \gg \ev{n_\darkmed}$.

The distribution of the energy $E_\darkmed$ of radiated $\darkmed$ bosons can
be obtained in a similar way.  With the notation $Z = E_\darkmed / E_0$, it is
given by
\begin{align}
  f_\darkmed(Z) = \frac{1}{\ev{n_\darkmed}} \sum_{m=1}^\infty \sum_{k=1}^m p_m \,
              f_{\darkmed, k}(Z) \,,
\end{align}
where $f_{\darkmed, k}(Z)$ is the energy distribution of the $k$-th emitted
$\darkmed$ boson.  Note that $f_{\darkmed, k}(Z)$ is not a function of the total
number of radiated dark photons $m$ because each splitting process is independent.
The recursion relation for $f_{\darkmed, k}(Z)$ is
\begin{align}
  f_{\darkmed, k+1}(Z) = \int_{x_{\min}}^{x_{\max}} \! dx_k \, f_{\chi,1}(x_k)
    \frac{f_{\darkmed, k}(Z / x_k)}{x_k} \,
    \Theta(1-x_{\max} \le Z \le 1-x_{\min}) \,.
  \label{eq:recursion-A'}
\end{align}

\subsection{Energy spectra from an inverse Mellin transform}
\label{sec:ds-mellin}

A particularly elegant way of computing the energy spectrum $f_\chi(X)$ of dark
matter particles and the spectrum $f_\darkmed(Z)$ of dark photons is by first
computing the moments $\ev{X^s}$ and $\ev{Z^s}$ of these distributions, and
then applying an inverse Mellin transform to recover $f_\darkmed(Z)$ itself.
The Mellin transform of a function $f(X)$ is defined as~\cite{Arfken:2008}
\begin{align}
  \mathcal{M}[f](s) &\equiv \varphi(s) \equiv \int_0^\infty \!dX\, X^{s-1} f(X) \,,
  \label{eq:Mellin}
\end{align}
and the inverse transform is given by
\begin{align}
  f(X) = \frac{1}{2\pi i} \int^{c+i\infty}_{c-i\infty} \!ds\, X^{-s} \, \varphi(s) \,.
  \label{eq:Mellin-inverse}
\end{align}
The Mellin transform and its inverse are closely related to the Fourier and
Laplace transforms~\cite{Arfken:2008} and can therefore be evaluated
efficiently using the FFT (Fast Fourier Transform)
algorithm~\cite{Cooley:1965}.

For events in which a single $\darkmed$ is emitted, the moments of the DM
energy distribution after the emission, weighted by the probability of having
exactly one emission, are given by
\begin{align}
  p_1 \ev{X^s}_{1\darkmed}
   &= e^{-\ev{n_\darkmed}} \frac{\alpha_\darkmed}{2\pi}
      \int_{x_{\min}}^{x_{\max}} \! dx \, x^s \,
      \int_{t_{\min}}^{t_{\max}} \frac{dt}{t} P_{\chi\to\chi}(x) \nonumber\\[0.2cm]
   &\equiv e^{-\ev{n_\darkmed}} \ev{n_\darkmed} \overline{X^s} \,.
  \label{eq:ev-xs-1}
\end{align}
In the last line, we have defined
\begin{align}
  \overline{X^s} \equiv \frac{1}{\ev{n_\darkmed}} \frac{\alpha_\darkmed}{2\pi}
        \int_{x_{\min}}^{x_{\max}} \! dx \, x^s \,
        \int_{t_{\min}}^{t_{\max}} \frac{dt}{t} P_{\chi\to\chi}(x) \,.
  \label{eq:Xs-bar}
\end{align}
As in the previous section, we again take $x_{\min}$, $x_{\max}$,
$t_{\min}$ and $t_{\max}$ to have the same value for each splitting process,
and we neglect the $t$-dependence of $P_{\chi\to\chi}$.
If two $\darkmed$ bosons are emitted, we obtain analogously
\begin{align}
  p_2 \ev{X^s}_{2\darkmed}
    &= e^{-\ev{n_\darkmed}} \bigg( \frac{\alpha_\darkmed}{2\pi} \bigg)^2
       \int_{x_{\min}}^{x_{\max}} \! dx \, x^s \,
       \int_{t_{\min}}^{t_{\max}} \frac{dt}{t} \,
       \int_{x_{\min}}^{x_{\max}} \! dx' \, x'^s \,
       \int_{t_{\min}}^t \frac{dt'}{t'} \,
       P_{\chi\to\chi}(x) P_{\chi\to\chi}(x') \nonumber\\
    &\simeq e^{-\ev{n_\darkmed}} \frac{1}{2!} \bigg( \frac{\alpha_\darkmed}{2\pi} \bigg)^2
       \int_{x_{\min}}^{x_{\max}} \! dx \, x^s \,
       \int_{t_{\min}}^{t_{\max}} \frac{dt}{t} \,
       \int_{x_{\min}}^{x_{\max}} \! dx' \, x'^s \,
       \int_{t_{\min}}^{t_{\max}} \frac{dt'}{t'} \,
       P_{\chi\to\chi}(x) P_{\chi\to\chi}(x') \nonumber\\
       &= e^{-\ev{n_\darkmed}} \frac{\ev{n_\darkmed}^2}{2!} \overline{X^s}^2 \,.
    \label{eq:ev-xs-2}
\end{align}
In the second line, we have again used the approximations from sec.~\ref{sec:ds-kinematics}
to extend the integration region of the $t'$ integral (see the discussion below
eq.~\eqref{eq:p2}).  Eq.~\eqref{eq:ev-xs-2} can be easily generalized to the
case of $m$ emitted $\darkmed$ bosons:
\begin{align}
  p_m \ev{X^s}_{m \darkmed}
    &= e^{-\ev{n_\darkmed}} \frac{\ev{n_\darkmed}^m}{m!} \overline{X^s}^m\,.
  \label{eq:ev-vs-m}
\end{align}
Summing over $m$, we find the moments of the overall $\darkmed$ energy
distribution:
\begin{align}
  \varphi(s+1) \equiv \ev{X^s}
    = \sum_{m=0}^\infty p_m \, \ev{X^s}_{m\darkmed}
    = e^{-\ev{n_\darkmed}(1 - \overline{X^s})} \,.
  \label{eq:avg-xs}
\end{align}
An inverse Mellin transform of $\varphi(s)$ then yields the dark matter energy
spectrum $f_\chi(X)$. As can be seen from eq.~\eqref{eq:Mellin-inverse}, this
requires evaluating $\varphi(s)$ at complex values of $s$.  A numerically
stable way of doing this is by using the FFT algorithm to compute
$\overline{X^s}$ (eq.~\eqref{eq:Xs-bar}), then using eq.~\eqref{eq:avg-xs}, and
afterwards apply another FFT to evaluate the inverse Mellin transform.

The procedure for obtaining the $\darkmed$ spectrum is alike and yields
for the contribution of showers with $m$ dark photons
to the moment $s$, weighted by the probability of emitting exactly $m$ dark photons,
\begin{align}
  p_m \ev{Z^s}_{m \darkmed}
    &= \frac{1}{\ev{n_\darkmed}} e^{-\ev{n_\darkmed}} \frac{\ev{n_\darkmed}^m}{m!}
       \overline{Z^s} \sum_{k=1}^m \overline{X^s}^{k-1} \,,
\end{align}
with
\begin{align}
  \overline{Z^s}
    &\equiv \frac{1}{\ev{n_\darkmed}} \frac{\alpha_\darkmed}{2\pi}
      \int_{x_{\min}}^{x_{\max}} \! dx \, (1 - x)^s \,
      \int_{t_{\min}}^{t_{\max}} \frac{dt}{t} P_{\chi\to\chi}(x) \,.
\end{align}
Summing over all emissions we obtain
\begin{align}
  \ev{Z^s} = \frac{\overline{Z^s}}{\ev{n_\darkmed}}
    \frac{1 - e^{-\ev{n_\darkmed}(1 - \overline{X^s})}}{1 - \overline{X^s}} \,.
  \label{eq:avg-zs}
\end{align}
An inverse Mellin transform yields the $\darkmed$ spectrum $f_\darkmed(Z)$.

\subsection{Comparison of analytic results to Monte Carlo simulations}
\label{sec:ds-comparison}

We have also simulated the radiation of $\darkmed$ bosons fully numerically in
Pythia~8~\cite{Carloni:2010tw, Carloni:2011kk, Sjostrand:2014zea}, using the
``hidden valley'' model implemented therein.  Note that, like our analytic
calculations, also the dark photon shower implemented in Pythia includes only
dark photon radiation, $\chi \to \chi\darkmed$ and $\bar\chi \to \bar\chi \darkmed$,
not dark photon splitting, $\darkmed \to \bar\chi \chi$.

To compare the $\chi$ and $\darkmed$ energy spectra obtained using the
different approaches, we assume $\chi$ pair production at a center of mass
energy of $\sqrt{\hat{s}} = 1$~TeV. In Pythia, this is achieved by simulating
the process $e^+ e^- \to Z' \to \bar\chi \chi$, followed by the dark photon
shower. The results for several characteristic parameters of the dark parton
shower, namely $\ev{n_\darkmed}$, $\ev{X}$ and $\ev{Z}$, are shown in
table~\ref{tab:radiation}. Comparing Pythia results to the results obtained
using eqs.~\eqref{eq:ev-n}, eq.~\eqref{eq:avg-xs} (with $s=1$) and
eq.~\eqref{eq:avg-zs} (with $s=1$), we find excellent agreement.  We attribute
the remaining differences to the approximations we had to make in the
analytical calculation, in particular the assumption that the upper and lower
integration limits in $x$ and $t$ do not depend on $x$ and $t$ themselves.

\begin{table}
  \begin{ruledtabular}
  \begin{tabular}{ccccccccc}
      & $m_\chi$ & $m_\darkmed$ & $2\ev{n_\darkmed}$ & $2\ev{n_\darkmed}_\text{Pythia}$ & $\ev{X}$ & $\ev{X}_\text{Pythia}$ & $\ev{Z}$ & $\ev{Z}_\text{Pythia}$ \\
      & {[GeV]}  & [GeV]        &                    &                                  &          &                        &          &                        \\
    \hline
      & 50  & 1.5 & 2.130  & 2.340  & 0.873 & 0.837 & 0.119 & 0.140 \\
      & 50  & 0.4 & 2.848  & 3.084  & 0.871 & 0.835 & 0.091 & 0.107 \\
    A & 4   & 1.5 & 3.476  & 3.540  & 0.729 & 0.697 & 0.156 & 0.171 \\
      & 4   & 0.4 & 4.990  & 4.825  & 0.712 & 0.681 & 0.116 & 0.132 \\
    B & 0.4 & 0.4 & 5.691  & 5.215  & 0.626 & 0.608 & 0.132 & 0.150 \\
  \end{tabular}
  \end{ruledtabular}
  \caption{Characteristics of the dark photon shower for various choices of
    the DM mass $m_\chi$ and the dark photon mass $m_\darkmed$. The rows
    labeled ``A'' and ``B'' correspond to the two benchmark points from
    table~\ref{tab:Benchmark}.  In all cases, we assume $\chi$ pair production
    at a center of mass energy $\sqrt{\hat{s}} = 1$~TeV and we take
    $\alpha_\darkmed = 0.2$.  We show the predicted number $2\ev{n_\darkmed}$
    of dark photons per event (with the factor of 2 coming from the fact that
    we consider $\chi$ pair production), the average fraction $\ev{X}$ that a
    $\chi$ particle retains of its initial energy after showering, and the
    average fraction $\ev{Z}$ of the initial $\chi$ energy that each dark
    photon receives.  We compare the predictions of our semi-analytic
    calculations, eqs.~\eqref{eq:ev-n}, \eqref{eq:avg-xs} and \eqref{eq:avg-zs}
    (with $s=1$), to the results of a Monte Carlo simulation in Pythia.  As
    expected, the results satisfy the relation $\ev{X} + \ev{n_\darkmed} \ev{Z}
  = 1$.}
  \label{tab:radiation}
\end{table}

\begin{figure}[t]
  \centering
  \includegraphics[width=0.45\textwidth]{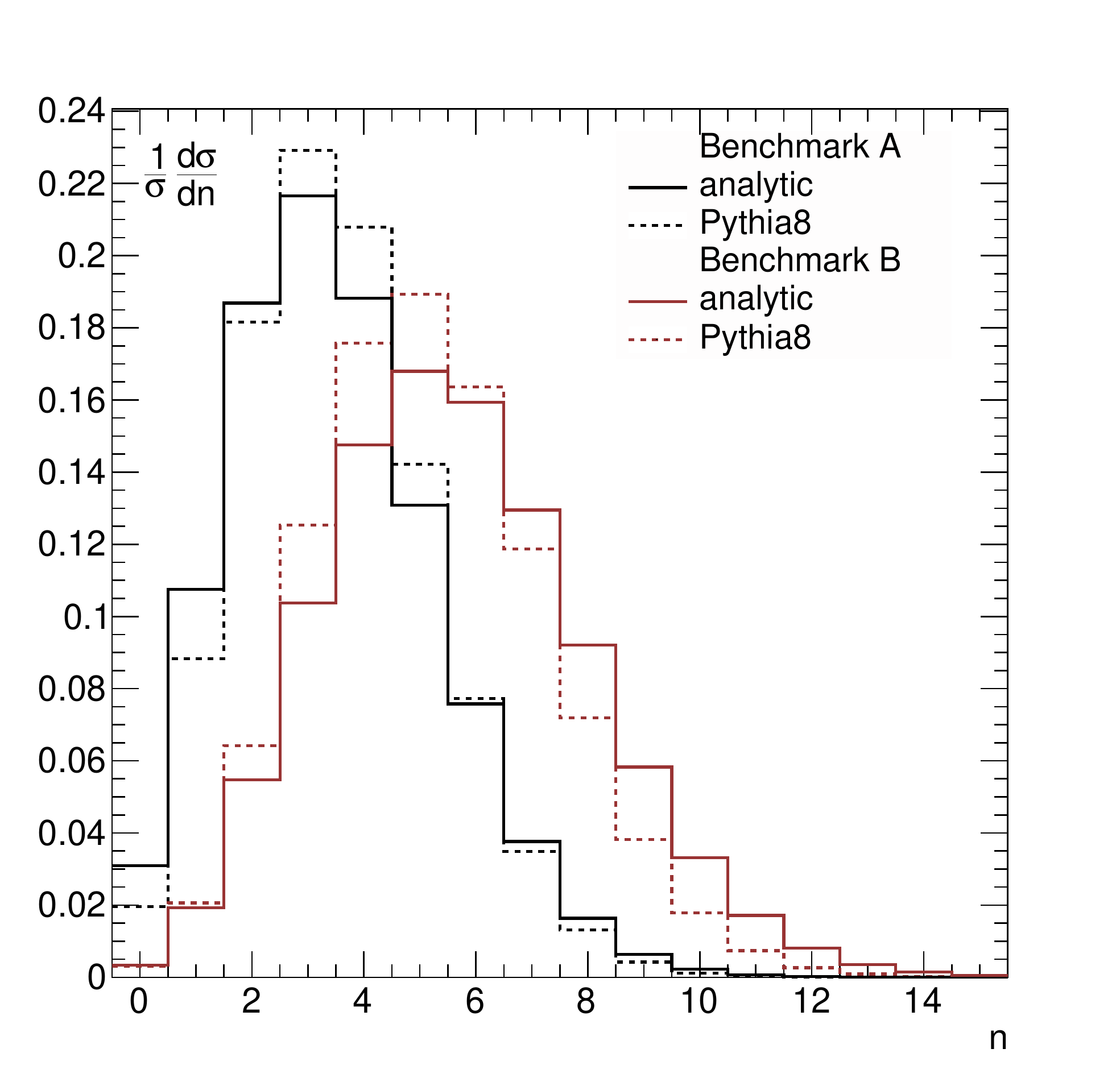}
  \caption{The distribution of the number $n$ of dark photons
    emitted in each $\bar\chi\chi$ pair production event at a center of
    mass energy of $\sqrt{\hat{s}} = 1$~TeV.  The model parameters are given
    in table~\ref{tab:Benchmark}. The solid curves labeled ``analytic'' show the
    Poisson probability $e^{-2\ev{n_\darkmed}} [2 \ev{n_\darkmed}]^n / n!$,
    with $\ev{n_\darkmed}$ given by eq.~\eqref{eq:ev-n}. The factors
    of 2 arise from the fact that two DM particles are produced in each event.
    The dotted curves show the distribution obtained from a Monte Carlo
    simulation of the dark photon shower in Pythia.}
  \label{fig:radSpec}
\end{figure}

\begin{figure}
  \begin{tabular}{cc}
    \includegraphics[width=0.45\textwidth]{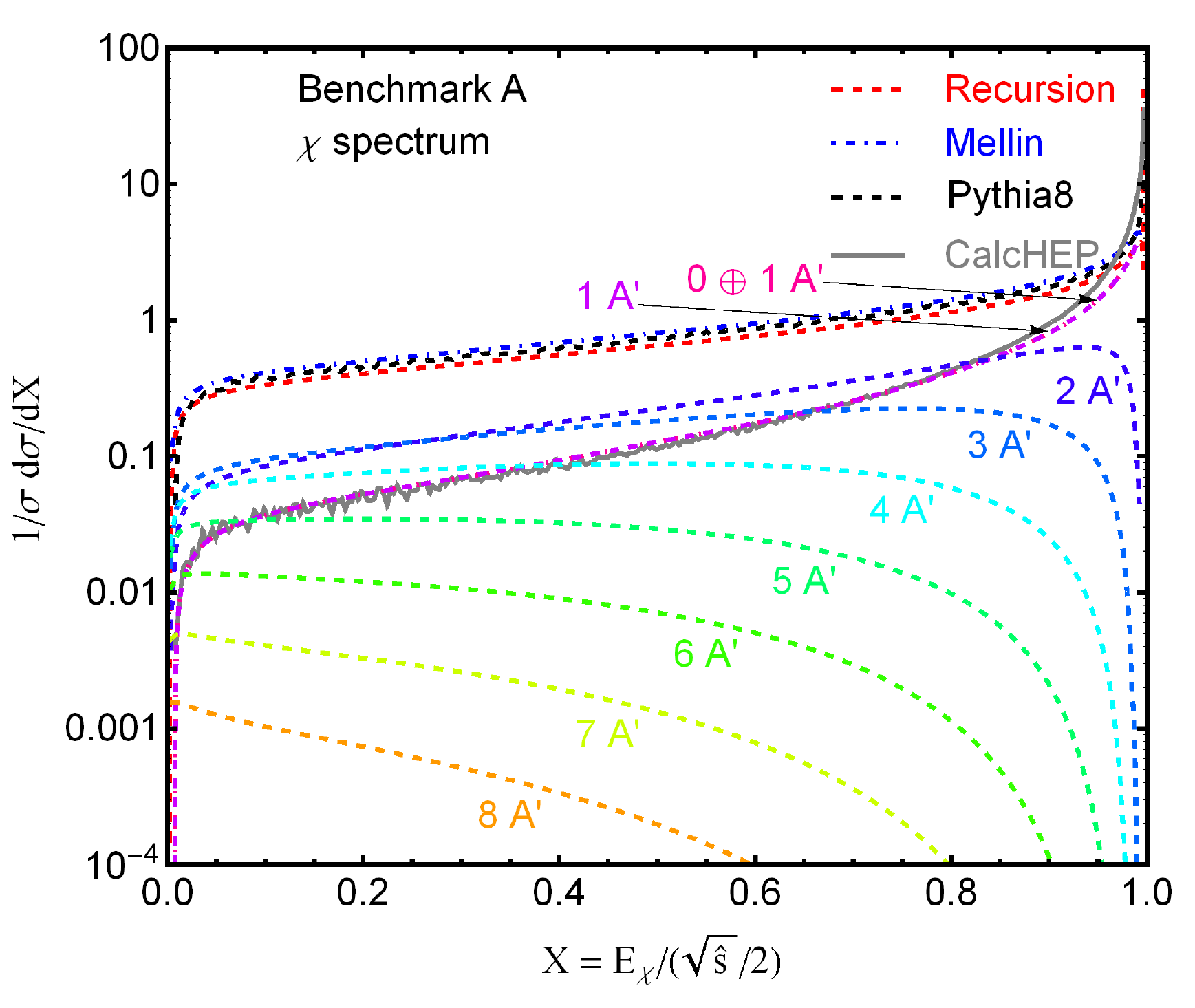} &
    \includegraphics[width=0.45\textwidth]{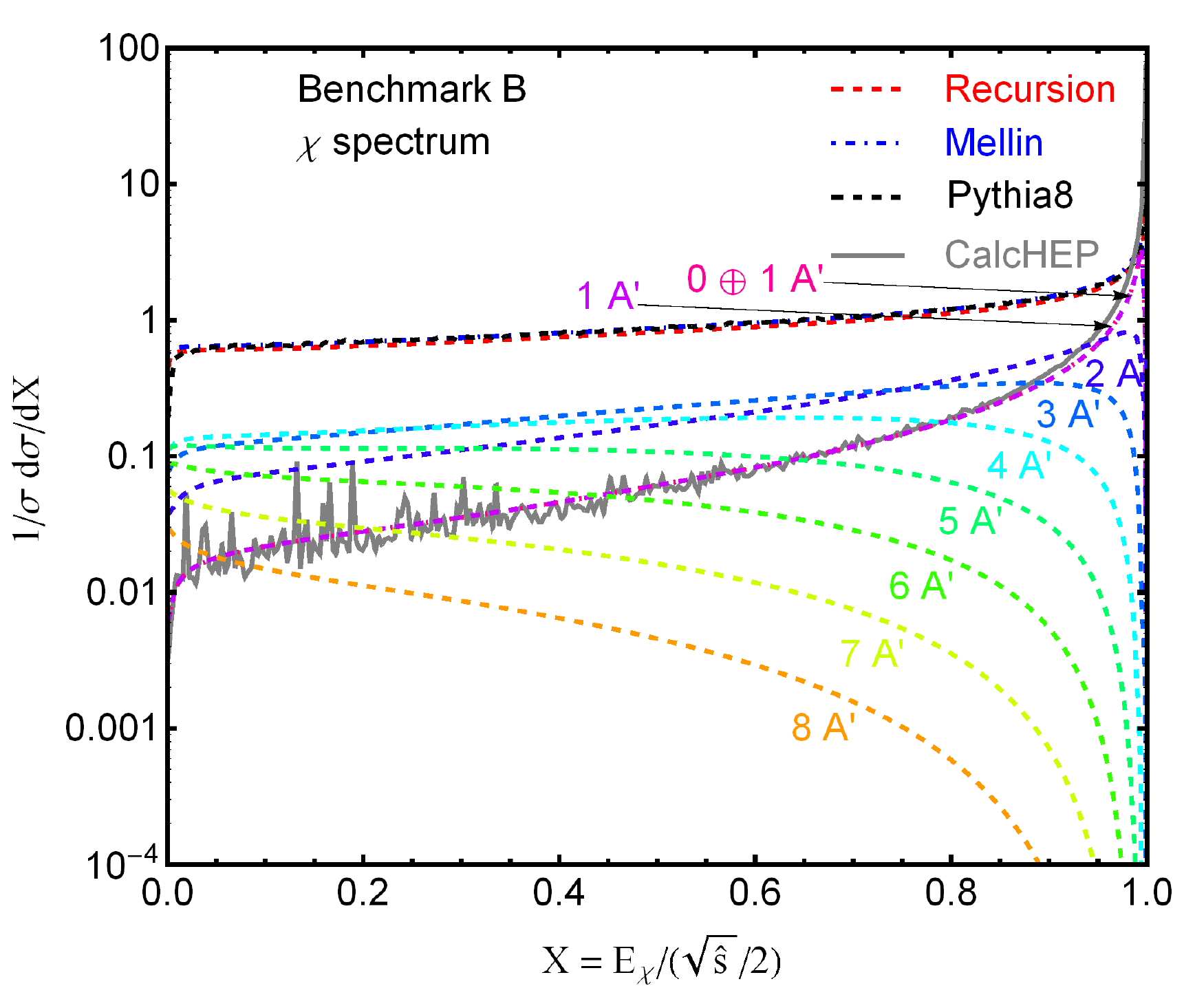} \\
    (a) & (b) \\
    \includegraphics[width=0.45\textwidth]{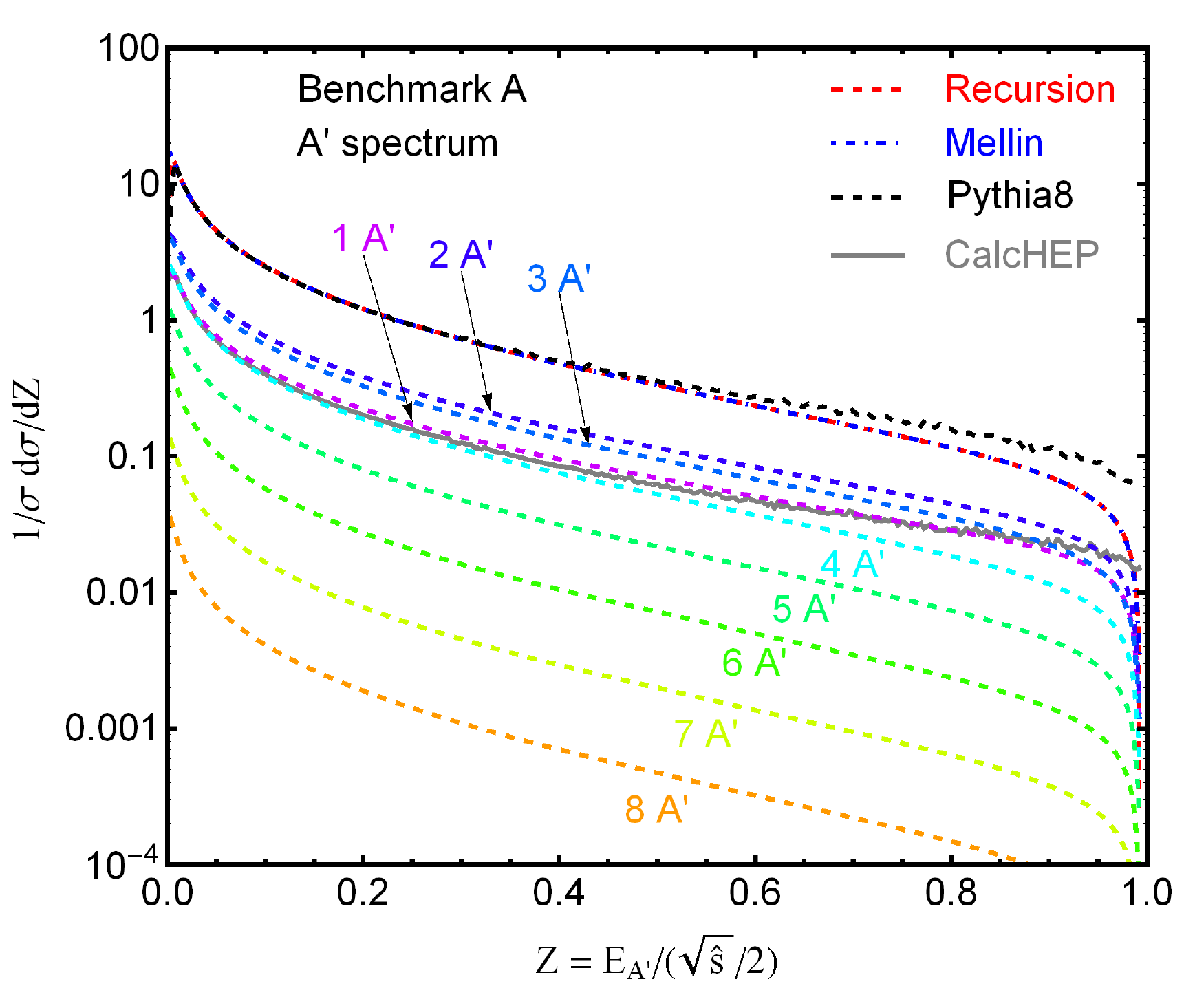} &
    \includegraphics[width=0.45\textwidth]{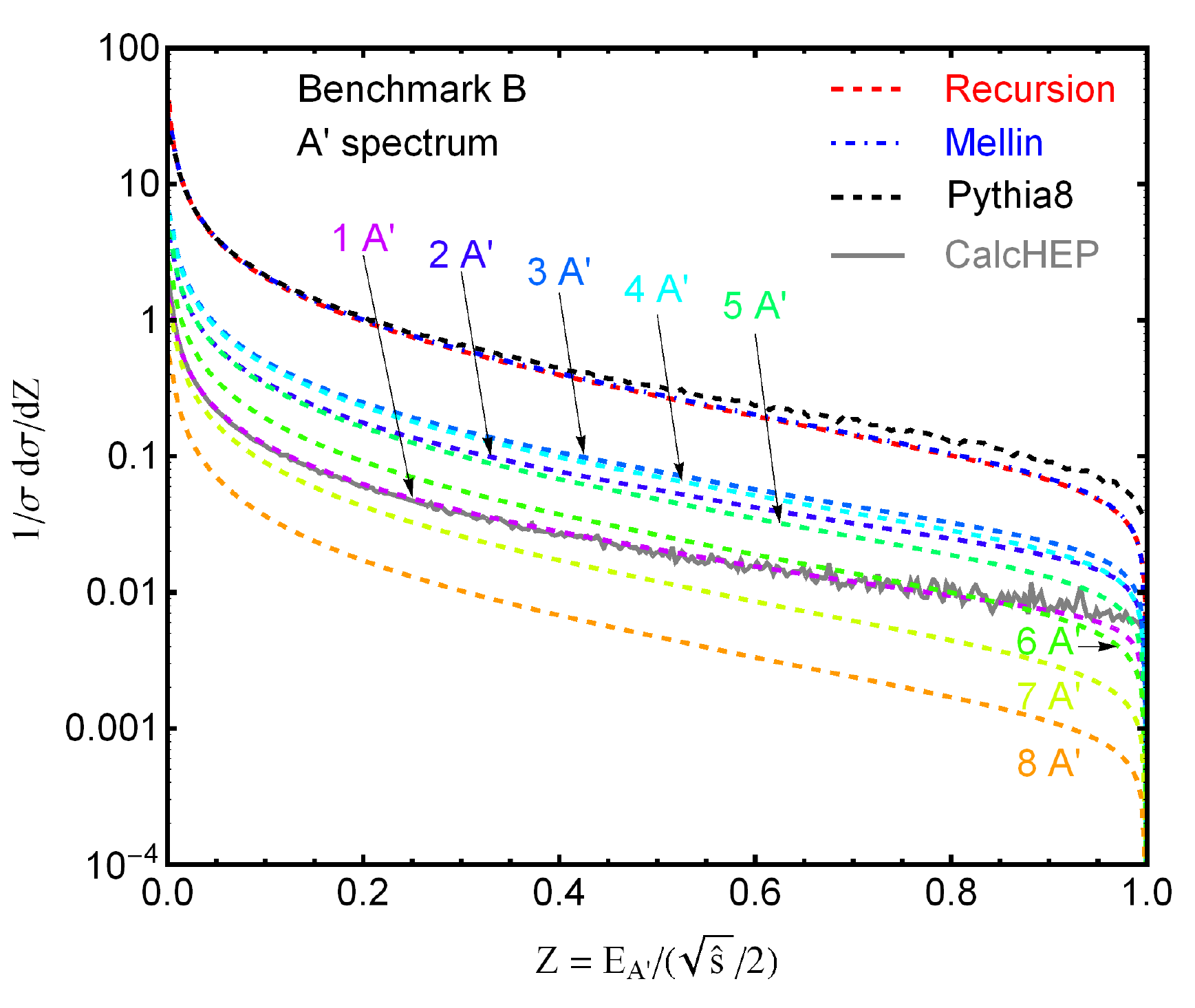} \\
    (c) & (d)
  \end{tabular}
  \caption{(a), (b) Energy spectrum $f_\chi(X)$ of DM particles $\chi$ after
    final state radiation; (c), (d) energy spectrum $f_\darkmed(Z)$ of dark
    photons $\darkmed$ emitted as final state radiation. The panels on the left
    are for benchmark point A from table~\ref{tab:Benchmark}, the panels on the
    right are for benchmark point B.  In all cases, we assume $\chi$ pair
    production at a center of mass energy $\sqrt{\hat{s}} = 1$~TeV.
    We compare the results from the recursion formulas described in
    sec.~\ref{sec:ds-recursion}, the Mellin transform method developed in
    sec.~\ref{sec:ds-mellin}, the dark photon shower simulation in Pythia,
    and a simple leading order simulation of $e^+ e^- \to \heavymed \to
    \bar\chi \chi \darkmed$ in CalcHEP.  For the Mellin transform method,
    we also show the result separated according to the number
    of $\darkmed$ bosons emitted in each $\bar\chi \chi$ pair production event.}
  \label{fig:ApDMSpec}
\end{figure}

A more detailed comparison between our analytic results and Pythia is shown in
figs.~\ref{fig:radSpec} and \ref{fig:ApDMSpec}. In fig.~\ref{fig:radSpec}, we
have plotted the distribution of the number of emitted dark
photons per event (i.e.\ per $\bar\chi \chi$ pair) at our two benchmark points
from table~\ref{tab:Benchmark}.  Fig.~\ref{fig:ApDMSpec} shows the energy
spectra of $\chi$ and $\darkmed$ after the dark photon shower has developed.

As we can see in fig.~\ref{fig:radSpec}, Pythia results confirm that dark photon
emission is approximately a Poisson process. We attribute the small deviations
to the approximations we made in deriving eq.~\eqref{eq:pm} in sec.~\ref{sec:ds-n}.

Regarding the energy spectra plotted in fig.~\ref{fig:ApDMSpec}, we find
excellent agreement between both of our analytic approaches (red dashed and
blue dot-dashed curves) and also with Pythia results (black dashed curves). Note
that the recursive method is numerically less stable than the Mellin transform
approach because numerical errors---especially those incurred close to the upper
integration boundary $x_{\max}$ at which the splitting kernel has a regularized
singularity---can accumulate in the recursive formulas \eqref{eq:recursion-chi}
and \eqref{eq:recursion-A'}.  We have also carried out simple leading order
simulations of the processes $e^+ e^- \to \heavymed \to \bar\chi \chi$ and
$e^+ e^- \to \heavymed \to \bar\chi \chi \darkmed$ in
CalcHEP~\cite{Belyaev:2012qa} (gray solid curves in fig.~\ref{fig:ApDMSpec}) and
find that, as expected, this simulation reproduces our analytic result for the
specific process $e^+ e^- \to \heavymed \to \bar\chi \chi \darkmed$ very well,
but is insufficient as a proxy for the full dark photon spectrum.  Since the
full dependence of the matrix element on $m_\chi$ and $m_\darkmed$ is taken
into account in CalcHEP, it also illustrates that the approximations discussed
in sec.~\ref{sec:ds-kinematics}
are justified at this level. In fig.~\ref{fig:ApDMSpec}, we also show the contributions
$p_m f_{\chi,m}(X)$ and $(1/\ev{n_\darkmed}) \sum_{k=1}^m p_m f_{\darkmed,k}(Z)$ from
events with a particular fixed number of dark photons to the overall energy
spectra $f_\chi(X)$ and $f_\darkmed(Z)$ (dashed rainbow-colored curves). Note
that events with several emitted $\darkmed$ bosons can dominate over events
with a single $\darkmed$ because $2\ev{n_\darkmed} > 1$ at our benchmark points.
This illustrates that the $\darkmed$ spectrum would be very difficult to obtain
from a full matrix element calculation and necessitates a parton shower
algorithm.

In summary, we have developed in this section two analytic methods for
computing the energy spectrum of $\chi$ and $\darkmed$, and we have verified
them by comparing to Monte Carlo simulations.

\section{Collider Searches for Lepton Jets from Dark Radiation}
\label{sec:collider}

To constrain radiating DM models using existing LHC lepton jet searches, we
numerically simulated the expected signal and background event spectra at
center of mass energies of $\sqrt{s} = 7$~TeV and $\sqrt{s} = 8$~TeV.
As in sec.~\ref{sec:ds-comparison}, signal events are generated using the
implementation of a ``hidden valley'' model in Pythia~8~\cite{Carloni:2010tw,
Carloni:2011kk, Sjostrand:2014zea}.  We compare our simulation results
to data from a 7~TeV ATLAS search for prompt lepton jets~\cite{Aad:2012qua} and
from an 8~TeV ATLAS search for displaced lepton jets~\cite{Aad:2014yea}.
We have also simulated events at $\sqrt{s} = 13$~TeV to predict the sensitivity
of future searches for prompt and displaced lepton jets.

\subsection{Prompt lepton jets}
\label{sec:prompt}

Our prompt lepton jet analysis follows the 7~TeV, 5~fb$^{-1}$ ATLAS search in
ref.~\cite{Aad:2012qua}, but including only muonic lepton jets.  The reason for
discarding lepton jets with electrons is that it is very difficult to implement
the corresponding selection criteria without a full detector simulation.  The
trigger requirement for muonic lepton jets is the presence of either three
muons with transverse momenta $p_T > 6$~GeV or one muon with $p_T > 18$~GeV.
Muons are required to have pseudorapidity $|\eta| < 2.5$, and they are required
to have a track in the inner detector associated with them.  We implement the
last requirement by demanding that the parent $\darkmed$ of the muon decays
within a radius of 122.5~mm (corresponding to the position of the last silicon
pixel layer) from the beam axis.  Finally, muon tracks are required to have a
transverse impact parameter $|d_0| < 1$~mm with respect to the primary vertex.

We cluster muons into lepton jets using the following algorithm: we collect all
muons with a cone of radius $\Delta R = 0.1$ around the highest-$p_T$ muon in
the event to form the first lepton jet. We then iterate this procedure,
starting, in each iteration, with the highest-$p_T$ muon not associated with a
lepton jet yet.  Ref.~\cite{Aad:2012qua} then defines two classes of events
that are considered in the analysis:

\emph{Double muon jet events} require
at least two muonic lepton jets, each of which must contain at least two
muons with $p_T > 11$~GeV. If the event was selected only by the single
muon trigger, the leading muon in the lepton jet is in addition required
to have $p_T > 23$~GeV.  Moreover, the two
muons closest in $p_T$ within each lepton jet are required to have an
invariant mass $m_{\mu\mu} < 2$~GeV.  To reject non-isolated lepton jets,
the variable
\begin{align}
  \rho \equiv \frac{\sum_i E_{T,i}}{p_{T,\text{LJ}}} \,,
  \label{eq:isolation-prompt}
\end{align}
is defined, where the sum in  the numerator runs over the $E_T$ of all
calorimeter deposits within $\Delta R = 0.3$
of any muon in the lepton jet, but excluding deposits within $\Delta R < 0.05$
around any muon. The denominator is the $p_T$ of the lepton jet.
The isolation cut imposes $\rho < 0.3$.

\emph{Single muon jet events} require only one muonic lepton jet, but this
lepton jet must contain at least four muons with $p_T > 19$~GeV, $16$~GeV,
$14$~GeV for the three leading muons and $p_T > 4$~GeV for all other muons.
The same invariant mass cut as for the double muon jet analysis is imposed,
and the isolation variable \eqref{eq:isolation-prompt} has to satisfy
$\rho < 0.15$.

Since we find that double muon jet events are much more sensitive to our
radiating DM model than single muon jet events, we do not include the latter
in the following analysis.

The dominant background to the muonic lepton jet search is from misidentified
QCD multijet events, which contribute $0.5 \pm 0.3$~events in the signal
region~\cite{Aad:2012qua}.  For the 7~TeV analysis, we can use this number at
face value, while for the extrapolation to $\sqrt{s} = 13$~TeV, we assume the
background cross section after cuts to be a factor of 3 larger.  This factor of
3 corresponds roughly to the scaling of the cross section for QCD jet
production~\cite{Campbell:2015qma}.  We assume the relative error on the
background to remain the same as at 7~TeV. Note that, to be able to use this
simple scaling of the background cross section, we have to use the same cuts at
7~TeV and at 13~TeV.

\subsection{Displaced lepton jets}
\label{sec:displaced}

Our search for displaced lepton jets, based on the ATLAS 8~TeV, 20.3~fb$^{-1}$
analysis~\cite{Aad:2014yea}, is somewhat
more involved than the search for prompt lepton jets described in the previous
section.  First, we need to be more careful in simulating detector effects,
which we do at particle level.  We pay particular attention to the displaced
$\darkmed$ decay vertices, which lead to unusual detector signatures for each
particle.  The coordinates of the $\darkmed$ decay vertex and the momenta of
the $\darkmed$ decay products are smeared by Gaussians with widths proportional
to $1/L_{xy}$, where $L_{xy}$ is the transverse distance of the $\darkmed$
decay vertex from the beam pipe.  This way, we in particular take into account
the fact that the momenta of charged particles that travel through more
material are affected more strongly by the smearing.  All parameters of our
detector simulation are calibrated against the lepton jet reconstruction
efficiencies shown in fig.~6 of ref.~\cite{Aad:2014yea}.  Since this plot is
for a sample of dark photons with flat $p_T$ and $\eta$ distributions in the
range $p_T\in[10,100]\ \text{GeV}$ and $\eta \in [-2.5,2.5]$, we generated a similar
event sample for the comparison.  To further improve the agreement between our
predicted efficiencies and the results from ref.~\cite{Aad:2014yea}, we apply a
fudge factor in each $L_{xy}$ bin.

Following~\cite{Aad:2014yea}, muons are selected from the event sample if they
are produced inside the sensitive ATLAS detector volume, i.e.\ up to a radius of
7~m.  Only loose requirements on the transverse momentum and pseudo-rapidity of
reconstructed muons are applied: their $p_T$ must be above the trigger
threshold of 6~GeV and the pseudorapidity $|\eta|$ must be $< 2.5$.  Since
the goal of the search in ref.~\cite{Aad:2014yea} was to constrain long-lived
dark photons with displaced decay vertices, the background from prompt muons is
reduced by requiring stand-alone muons, i.e.\ muons that are detected in the muon
spectrometer but cannot be matched to a track in the inner detector. This
essentially means that the dark photon decay in which the muon was produced
must have happened outside the last pixel layer of the inner detector (ID),
located at a radius of 122.5~mm from the beam pipe. To suppress cosmic ray
muons, the reconstructed muon trajectory must still be associated with the
primary interaction vertex.  In particular, it must fulfill requirements on the
impact parameters $|z_0| < 270$~mm (longitudinal) and $|d_0| < 200$~mm
(transverse).

Hadronic jets (or, more precisely, calorimeter jets) are based on energy
deposits in the electromagnetic calorimeters (ECAL) and hadronic calorimeters
(HCAL), clustered using the anti-$k_T$ algorithm \cite{Cacciari:2008gp}
implemented in FASTJET~\cite{Cacciari:2011ma} with a cone radius of $R = 0.4$.
Calorimeter jets must fulfill $p_T > 20~\GeV$ and $|\eta| < 2.5$.

Three different types of lepton jets (LJ) are defined: A type-0 or ``muonic''
LJ arises from dark photon showers in which all $\darkmed$ bosons decay to
muons. It consists of at least two displaced and collimated muons with no
additional activity nearby.  If however, such muons are accompanied by a single
calorimeter jet, we call the sum of these objects a type-1 or ``mixed'' LJ.  A
type-2 LJ or ``calorimeter'' LJ is defined as an isolated calorimeter jet that
fulfills additional requirements to distinguish it from QCD background.  Note
that, especially for type-1 and type-2 LJs, the term lepton jet used in
\cite{Aad:2014yea} is somewhat misleading because these objects can also arise
from hadronic $\darkmed$ decays. The clustering algorithm for all three LJ
types is described in the following.

We collect all muons and calorimeter jets inside a cone of radius $\Delta R =
0.5$ around the highest-$p_T$ muon in the event.  If at least one other muon
and no calorimeter jet is found inside this cone, those muons are combined to a
muonic LJ.  If at least one muon and exactly one calorimeter jet is found
inside the cone, all objects are combined into a mixed LJ.  If no second muon
is found within $\Delta R = 0.5$ of the highest-$p_T$ muon, the muon is
discarded.  The algorithm then continues with the highest-$p_T$ muon that has
not been associated with a lepton jet or discarded yet. After all muons have
been processed, the remaining calorimeter jets are reconstructed as calorimeter
LJs if their electromagnetic (EM) fraction is lower than 0.1.  This is for
example true for dark photons decaying to charged pions, but also for a decays
to electrons happening inside the hadronic calorimeter.  Since the
pseudo-rapidity region $1.0 < |\eta| < 1.4$ tends to create a fake low EM
fraction due to the transition between the barrel and endcap EM calorimeters
(ECAL), this region is excluded.  Additionally the width $W$ of a calorimeter
LJ, defined by
\begin{align}
  W \equiv \frac{\sum_i \Delta R^i \cdot p_T^i}{\Sigma_i p_T^i} < 0.1 \,,
\end{align}
needs to be small. Here, the sum runs over all particles in the lepton jet.  An
isolation cut is imposed on all reconstructed lepton jets by requiring that the
scalar $p_T$ sum of all charged tracks, reconstructed by the inner detector
inside a cone of radius 0.5 around the lepton jet axis, is below 3~GeV. Only
charged tracks with $p_T > 400$~MeV and with impact parameters $|z_0|<10$~mm
and $|d_0| < 10$~mm are considered in this procedure.  An event is discarded if there are
not exactly two lepton jets fulfilling $|\Delta\phi| > 1.0$. The requirement of
two lepton jets leads to the 6 different combinations 0--0, 0--1, 0--2, 1--1, 1--2,
and 2--2, where e.g.\ 0--1 corresponds to an event reconstructed with one type-0
(muonic) and one type-1 (mixed) lepton jet.

The sensitivity of the displaced analysis will depend strongly on the $\darkmed$ decay
mode.  Therefore, we list in table~\ref{tab:DetectorVolume} the most important
$\darkmed$ decay modes and their associated signatures. Check marks
($\checkmark$) indicate in which region of the detector a particular decay has
to happen in order to potentially be reconstructed as a displaced lepton jet.  For decays
that would fail our cuts, we give the reason why they are vetoed.  For
instance, due to the requirement of a low EM fraction, $\darkmed \to e^+ e^-$
is only identified as a displaced lepton jet if the decay happens inside the hadronic
calorimeter, while $\darkmed \to \mu^+ \mu^-$ is also selected if the decay
happens inside the electromagnetic calorimeter.  Decays to any final states
involving charged particles are vetoed if they happen in the inner detector.
Regarding decays to mesons, it is important to note that $\pi^\pm$, $K^\pm$ and
$K^0_L$ have lifetimes of order $10^{-8}$~sec~\cite{Agashe:2014kda}
and are thus stable on detector
scales (i.e.\ they will be stopped in the hadronic calorimeter before they
decay).  Neutral pions, on the other hand, decay very fast. For $K^0_S$, the lifetime of
$9 \times 10^{-11}$~sec may allow it to travel for several cm before decaying
(mostly to $\pi^+ \pi^-$ or $\pi^0 \pi^0$).  Therefore, $\darkmed \to \pi^+
\pi^-$, $K^+ K^-$ are accepted if they happen in the electromagnetic or
hadronic calorimeter. The decay $\darkmed \to \pi^+ \pi^- \pi^0$, on the other
hand is vetoed if it happens in the electromagnetic calorimeter because the
quasi-instantaneous decay of the $\pi^0$ to two photons deposits a large amount
of energy, violating the cut on the EM fraction.  The decay $\darkmed \to K^0_L
K^0_S$ is successfully reconstructed as a displaced lepton jet in many, but not all,
cases.  It will be vetoed in the inner detector if both the $\darkmed$ and the
secondary $K^0_S$ decay very fast (well within the 12.25~cm radius of the inner
detector) in spite of their typically large boosts, and the $K^0_S$ decay mode
is to $\pi^+ \pi^-$ (branching ratio 69\%).  It will be vetoed by the cut on
the EM fraction if the $\darkmed$ decay happens in the inner detector or the
electromagnetic calorimeter and the $K^0_S$ decay mode is to $\pi^0 \pi^0$
(branching ratio 31\%).

\begin{table}
  \begin{ruledtabular}
  \begin{tabular}{lccccc}
    Detector & $\darkmed\rightarrow e^+e^-$
      & $\darkmed\rightarrow \mu^+\mu^-$ & $\darkmed\rightarrow \pi^+\pi^- /K^+K^-$
      & $\darkmed\rightarrow \pi^+\pi^-\pi^0$
      & $\darkmed\rightarrow K^0_L K^0_S$\\\hline
    LJ type & 2 (calorimeter) & 0 (muonic) & 2 (calorimeter) & 2 (calorimeter)
                                                             & 2 (calorimeter) \\ \hline
    ID    & track       & track      & track      & track       & (\checkmark) \\
    ECAL  & EM fraction & \checkmark & \checkmark & EM fraction & (\checkmark) \\
    HCAL  & \checkmark  & \checkmark & \checkmark & \checkmark  & \checkmark
  \end{tabular}
  \end{ruledtabular}
  \caption{Illustration of where in the detector a specific $\darkmed$ decay
    must happen in order to potentially be reconstructed as a lepton jet in our
    displaced lepton jet analysis (see sec.~\ref{sec:displaced} and
    ref.~\cite{Aad:2014yea}.  For
    decays that will be vetoed, a reason for the veto is given. The type of
    lepton jet as which each decay mode is most likely to be reconstructed is given
    at the top of the table.}
  \label{tab:DetectorVolume}
\end{table}

Note that, as can be read from table~\ref{tab:DetectorVolume}, the decay mode
$\darkmed \to \mu^+ \mu^-$ is most likely to be reconstructed as a type-0
(muonic) lepton jet, while all other decay modes typically lead to type-2
(calorimeter) lepton jets.  Type-1 (mixed) lepton jets are obtained only if
several $\darkmed$ are radiated from the same dark matter particle,
with at least one of them decaying to $\mu^+ \mu^-$ and at
least one of them decaying to a different final state.  Therefore, the
requirement for a type-1 (mixed) lepton jet are the most difficult to satisfy.
The predicted signal event numbers in table.~\ref{tab:EventNumbers-displaced}
confirm this.

\subsection{Results}

In tables~\ref{tab:EventNumbers-prompt} and \ref{tab:EventNumbers-displaced}
we present the signal and background predictions as well as the observed event
numbers for the two benchmark models defined in table~\ref{tab:Benchmark}.
Table~\ref{tab:EventNumbers-prompt} is for the prompt lepton jet analysis
from ref.~\cite{Aad:2012qua}, using 5~fb$^{-1}$ of 7~TeV ATLAS data
(see sec.~\ref{sec:prompt}, while table~\ref{tab:EventNumbers-displaced} is for
the displaced lepton jet search from ref.~\cite{Aad:2014yea}, using 20.3~fb$^{-1}$
of 8~TeV ATLAS data (see sec.~\ref{sec:displaced}).
Note that, for the prompt search, we include only muonic lepton jets, which
are easier to model without a full detector simulation. For the displaced search,
we show event numbers with and without inclusion of the background-limited
sample of events with two type-2 lepton jets.
We see from tables~\ref{tab:EventNumbers-prompt} and \ref{tab:EventNumbers-displaced}
that, as expected, benchmark point~A is more easily detectable in the displaced
search, while benchmark point~B is probed more sensitively by the prompt search.

We also show our predictions for 100~fb$^{-1}$ of 13~TeV data. In
general, we expect better sensitivity at 13~TeV because of the larger
integrated luminosity expected.  Since an
estimation of the multijet background at 13~TeV is problematic, we restrict the
displaced lepton jet search at 13~TeV to type-0 (muonic) LJs, for which QCD
backgrounds do not play a major role.  The only relevant source of background
in the 13~TeV displaced search are then cosmic rays, whose flux is independent of
the collider center of mass energy as long as we use the same cuts as at
$\sqrt{s}=8$~TeV.  Our predictions for the sensitivity of a displaced lepton
jet search at 13~TeV LHC are thus very conservative.  For the prompt search at
13~TeV, we scale the background cross section from the 7~TeV analysis by a
factor of 3.

\begin{table}
  \begin{center}
    \begin{minipage}{10cm}
      \begin{ruledtabular}
      \begin{tabular}{lcc}
                        & 7 TeV         & 13 TeV        \\\hline
        Benchmark A     & 0.8           & 109           \\
        Benchmark B     & 3.9           & 334           \\
        All backgrounds & $0.5\pm0.3$   & $30\pm18$     \\
        data            & 3             &               \\
      \end{tabular}
      \end{ruledtabular}
    \end{minipage}
  \end{center}
  \vspace{-0.5cm}
  \caption{Predicted event rates for the prompt lepton jet analysis
    (see sec.~\ref{sec:prompt} and ref.~\cite{Aad:2012qua}) at the benchmark
    parameter points from table~\ref{tab:Benchmark}, compared to the background
    predictions and the observed event rates from ref.~\cite{Aad:2012qua}.}
  \label{tab:EventNumbers-prompt}
\end{table}

\begin{table}
  \begin{ruledtabular}
  \begin{tabular}{lcccccccc}
                          & \multicolumn{8}{c}{Lepton jet type}                                         \\\cline{2-9}
                          & 0-0 & 0-1 & 0-2 & 1-1 & 1-2 & 2-2 & All                & All excl. 2-2      \\\hline
    \quad Cosmic ray bkg. & 15  & 0   & 14  & 0   & 0   & 11  & $40\pm 11\pm 9$    & $29\pm 9\pm 29$    \\\hline
    {\bf 8~TeV}           &     &     &     &     &     &     &                    &                    \\
    \quad Multi-jet bkg.   &     &     &     &     &     &     & $70\pm 58\pm 11$   & $12\pm 9\pm 2$     \\
    \quad Benchmark A     & 14  & 3   & 104 & 0   & 14  & 200 & $335\pm 18\pm 100$ & $135\pm 12\pm 41$  \\
    \quad Benchmark B     & 2.1 & 0.4 & 3.0 & 0   & 0.3 & 1.2 & $7\pm 2.1\pm 2.6$  & $5.8\pm 1.7\pm 2.4$ \\
    \quad data            & 11  & 0   & 11  & 4   & 3   & 90  & 119                & 29                 \\\hline
    {\bf 13~TeV}          &     &     &     &     &     &     &                    &                    \\
    \quad Benchmark A     & 169 &   &  &    &   &  &  & \\
    \quad Benchmark B     & 28  &   &  &    &   &  &  &
  \end{tabular}

  \end{ruledtabular}
  \caption{Predicted event rates for the displaced lepton jet analysis
    (see sec.~\ref{sec:displaced} and ref.~\cite{Aad:2014yea})
    at the benchmark parameter points
    from table~\ref{tab:Benchmark},
    compared to the background predictions and the observed event rates
    from ref.~\cite{Aad:2014yea}. In the last two columns, the
    first error is the statistical uncertainty, while the second one
    is systematic. Our sensitivity study at $\sqrt{s} = 13$~TeV includes
    only type 0--0 events (only muonic lepton jets) because a reliable
    extrapolation of the multijet background to 13~TeV is difficult.}
  \label{tab:EventNumbers-displaced}
\end{table}

\begin{figure}
  \vspace{-0.9cm}
  \begin{center}
  \begin{tabular}{cc}
    \includegraphics[width=0.45\textwidth]{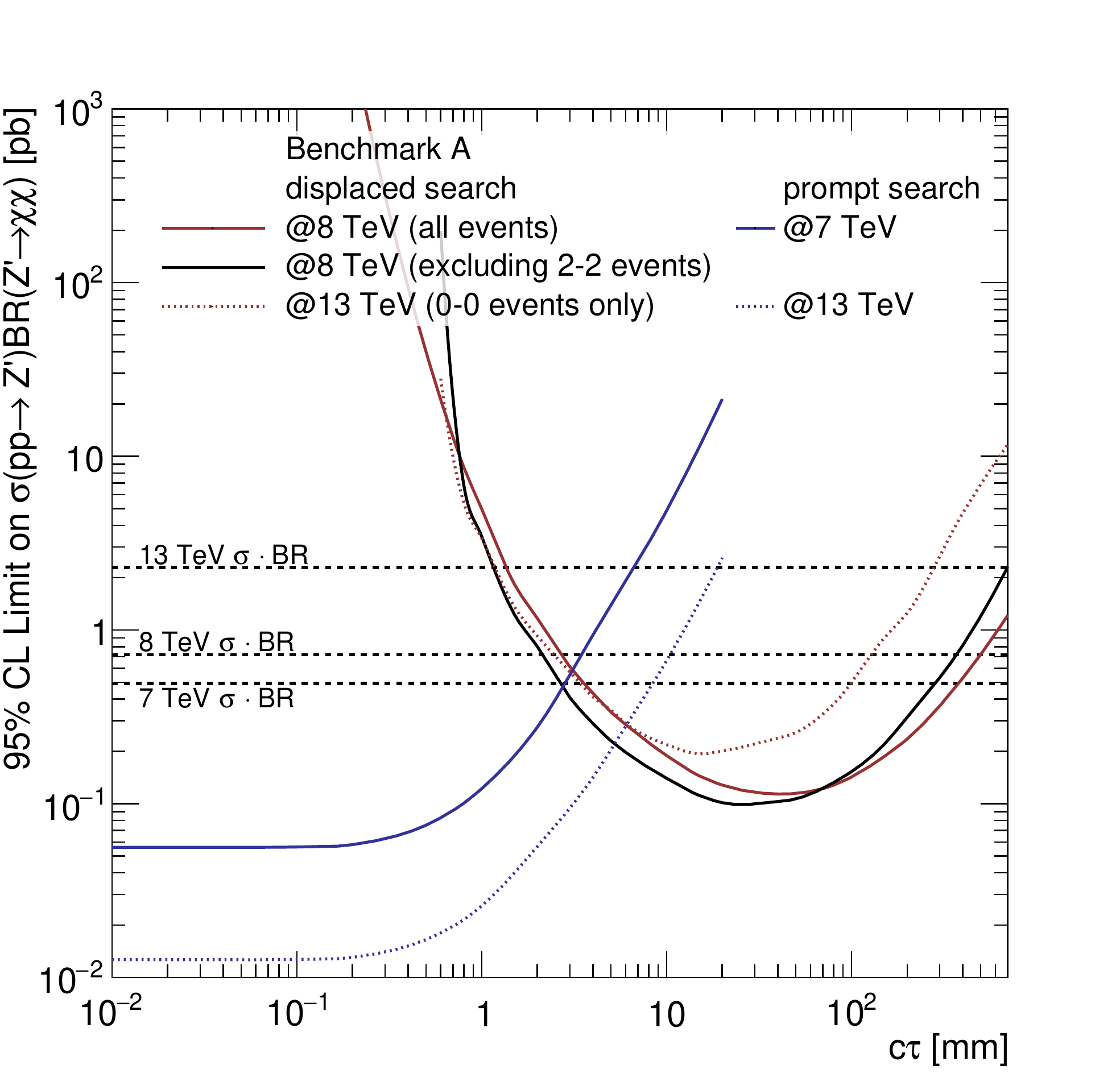} &
    \includegraphics[width=0.45\textwidth]{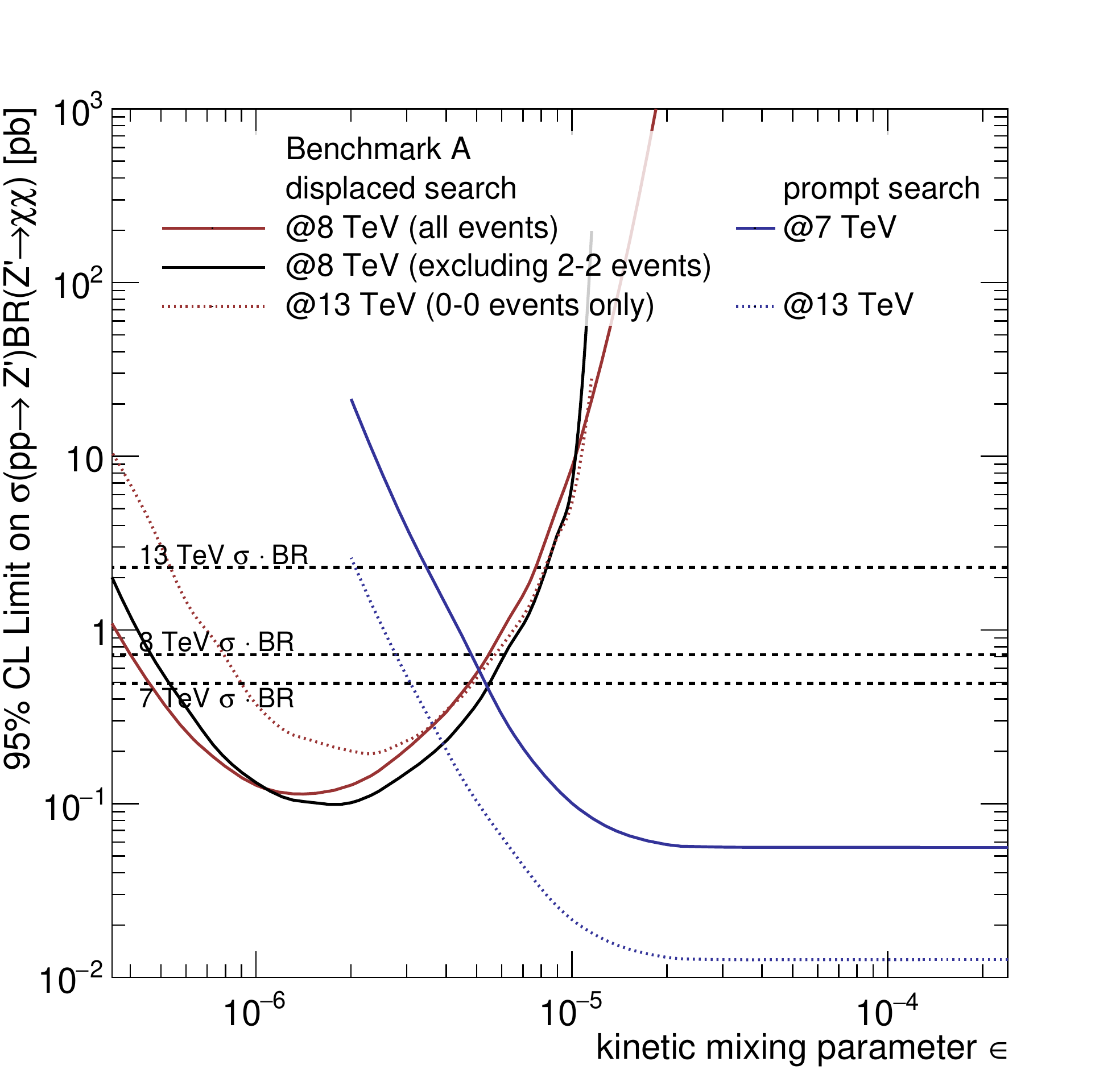} \\[-0.3cm]
    (a) & (b) \\[-0.4cm]
    \includegraphics[width=0.45\textwidth]{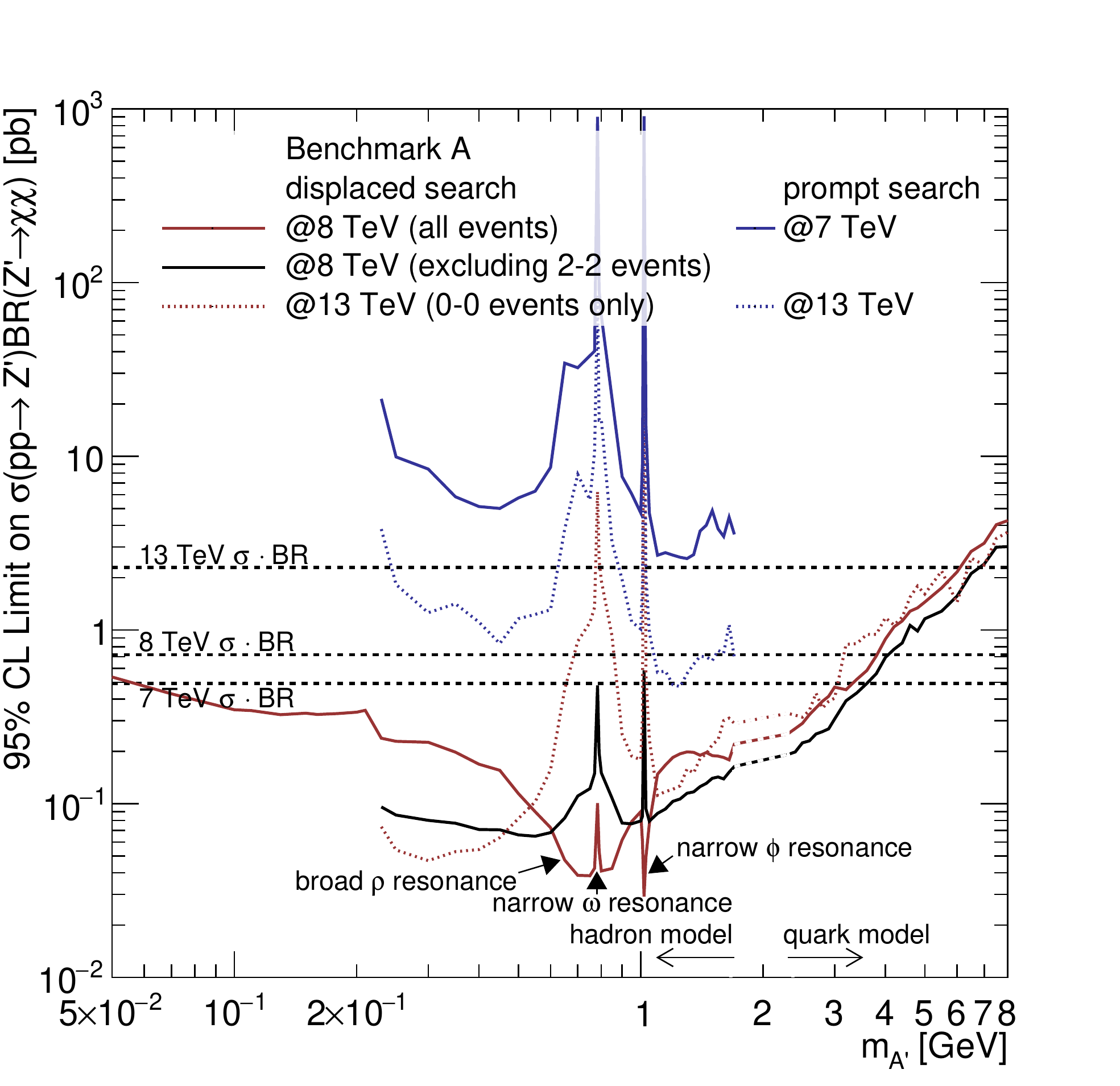} &
    \includegraphics[width=0.45\textwidth]{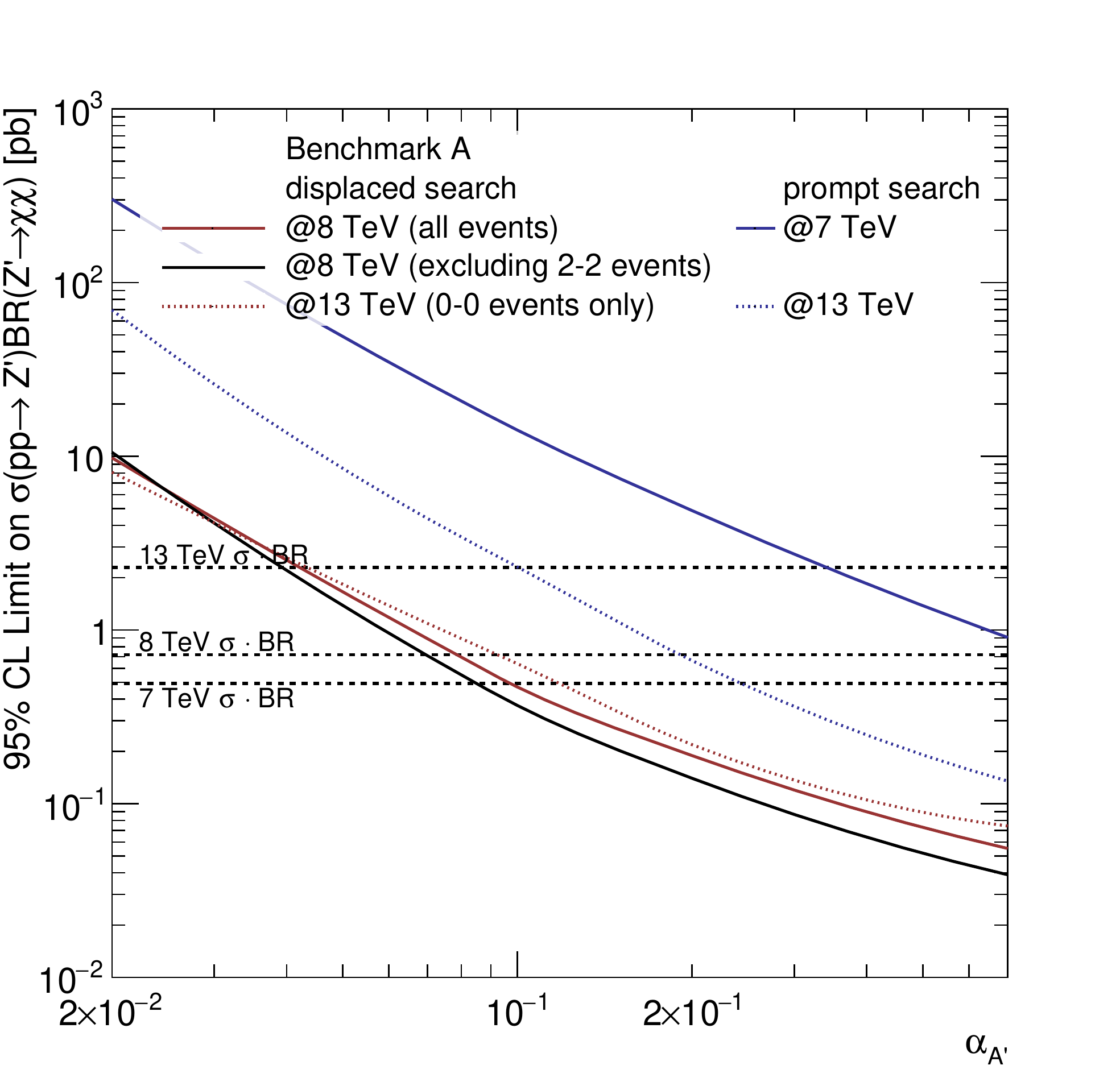} \\[-0.3cm]
    (c) & (d) \\[-0.4cm]
    \includegraphics[width=0.45\textwidth]{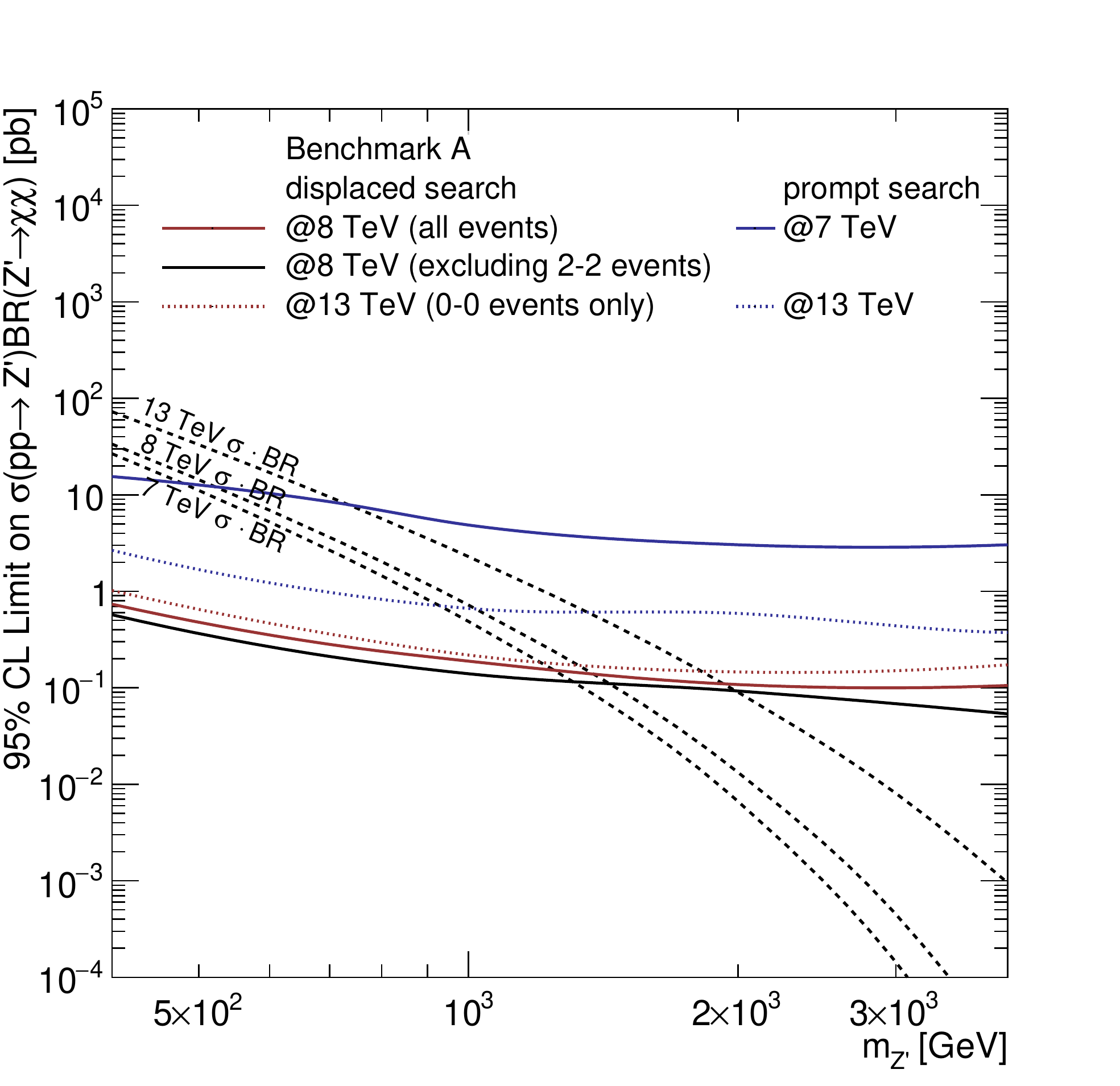} &
    \includegraphics[width=0.45\textwidth]{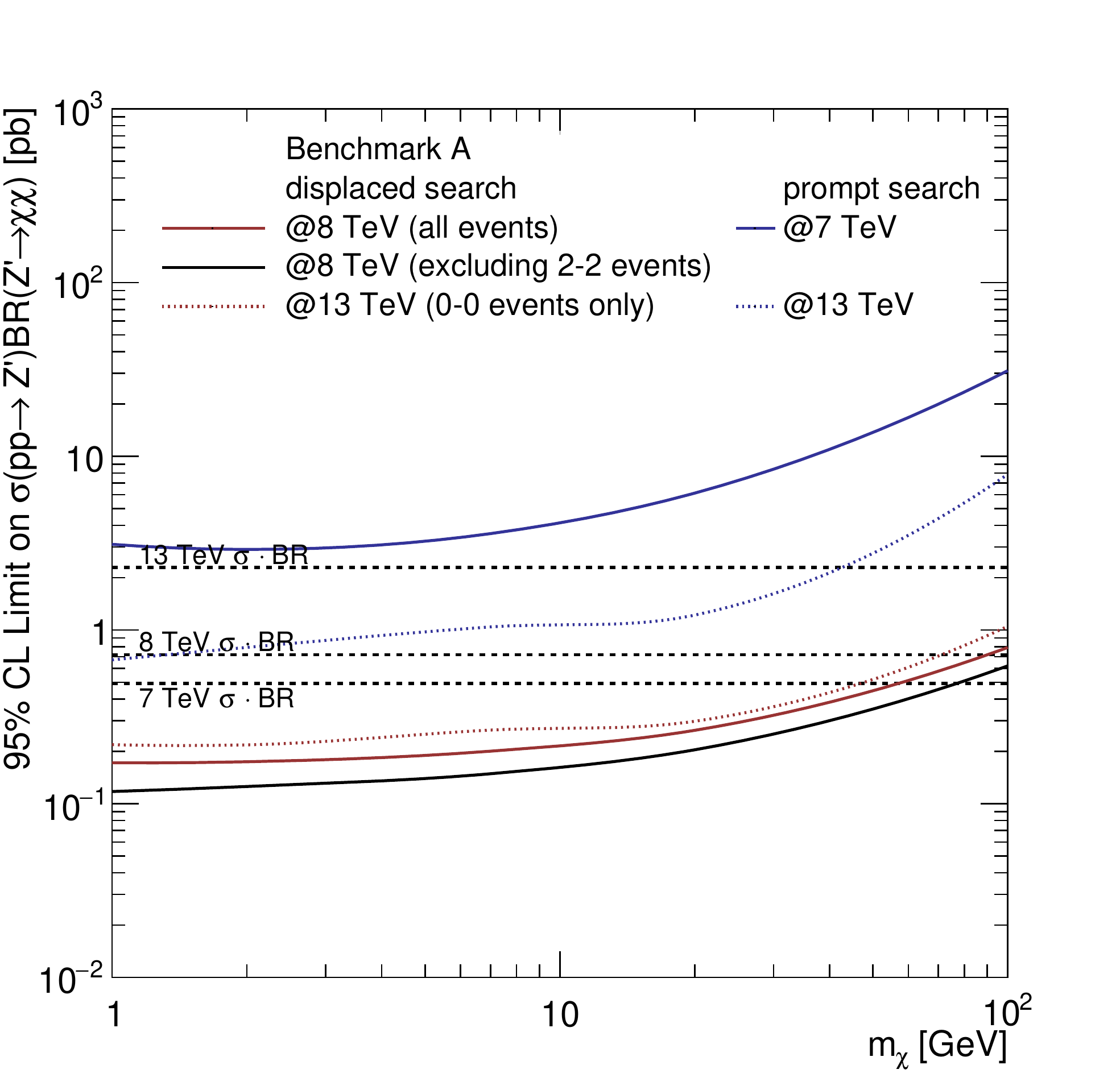} \\[-0.3cm]
    (e) & (f)
  \end{tabular}
  \end{center}
  \vspace{-0.5cm}
  \caption{95\% CL upper limits on the $\text{DM}\;\text{DM} + n \darkmed$
    production cross section as a function of the model parameters for
    benchmark point~A from table~\ref{tab:Benchmark}.  In each panel, we vary
    one parameter while keeping the others fixed at their benchmark values.
    Exclusion limits from the 7~TeV ATLAS search for prompt lepton
    jets~\cite{Aad:2012qua} (solid blue) and from the 8~TeV ATLAS search for
    displaced lepton jets~\cite{Aad:2014yea} are shown. For the latter search,
    we show results including all lepton jet events (red solid) and excluding
    events with two type-2 lepton jets (black solid).  The predicted
    sensitivity of similar analyses at $\sqrt{s} = 13$~TeV is shown as blue/red
    dotted curves.  The black dotted lines in each panel show the theoretically
    predicted production cross sections.}
  \vspace*{-0.5cm}
  \label{fig:CLs1}
\end{figure}

\begin{figure}
  \vspace{-0.9cm}
  \begin{center}
  \begin{tabular}{cc}
    \includegraphics[width=0.45\textwidth]{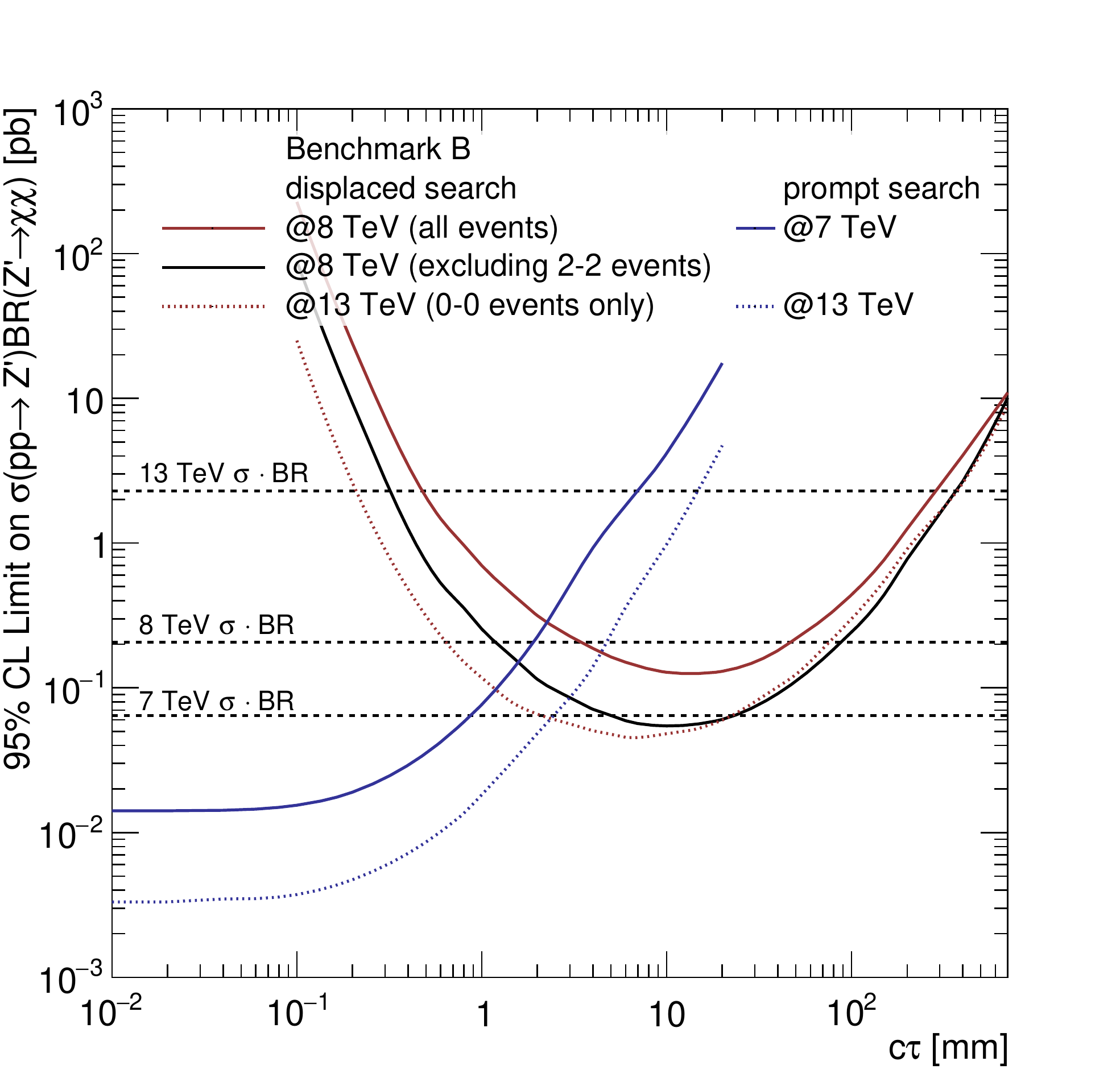} &
    \includegraphics[width=0.45\textwidth]{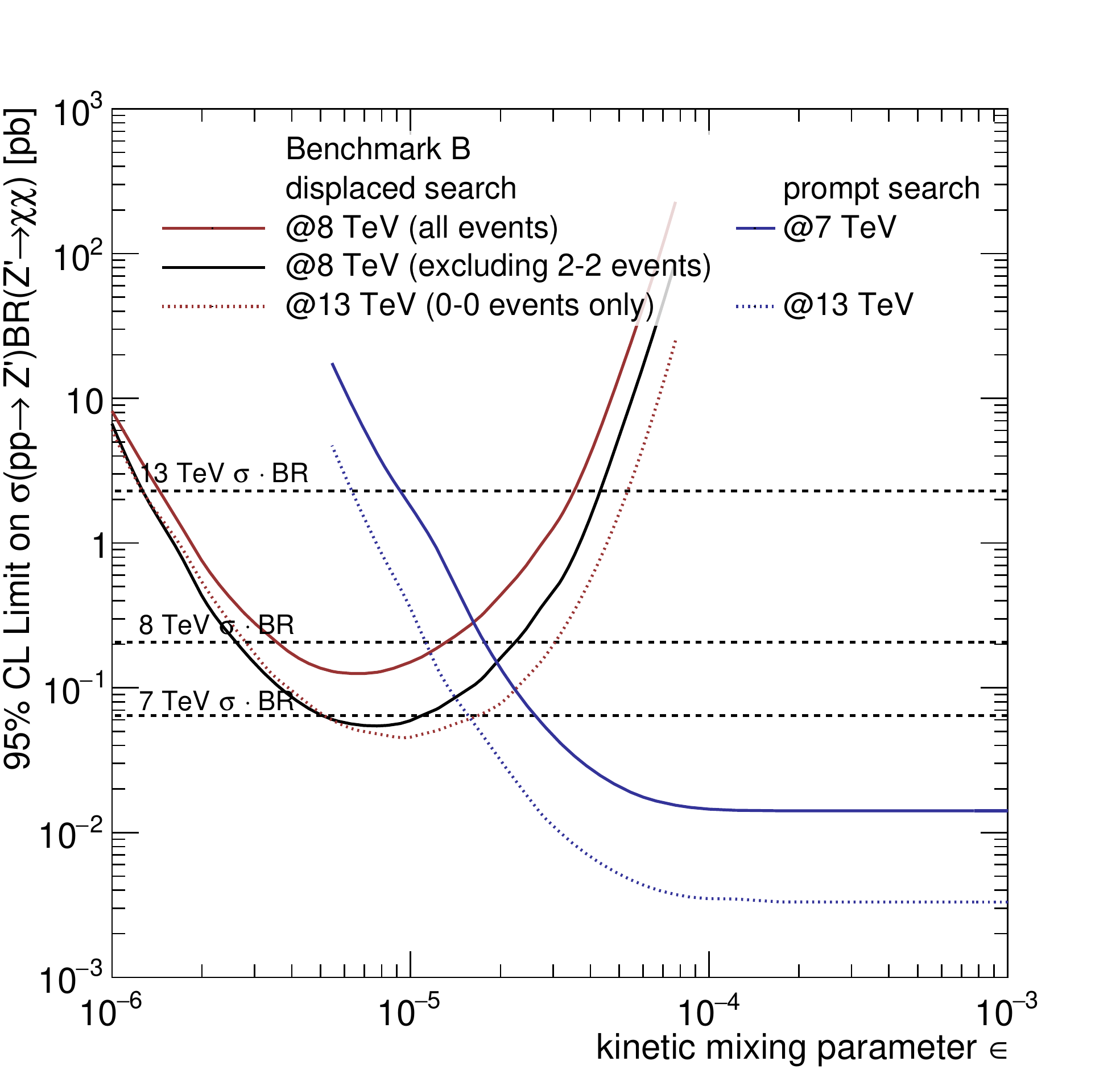} \\[-0.3cm]
    (a) & (b) \\[-0.4cm]
    \includegraphics[width=0.45\textwidth]{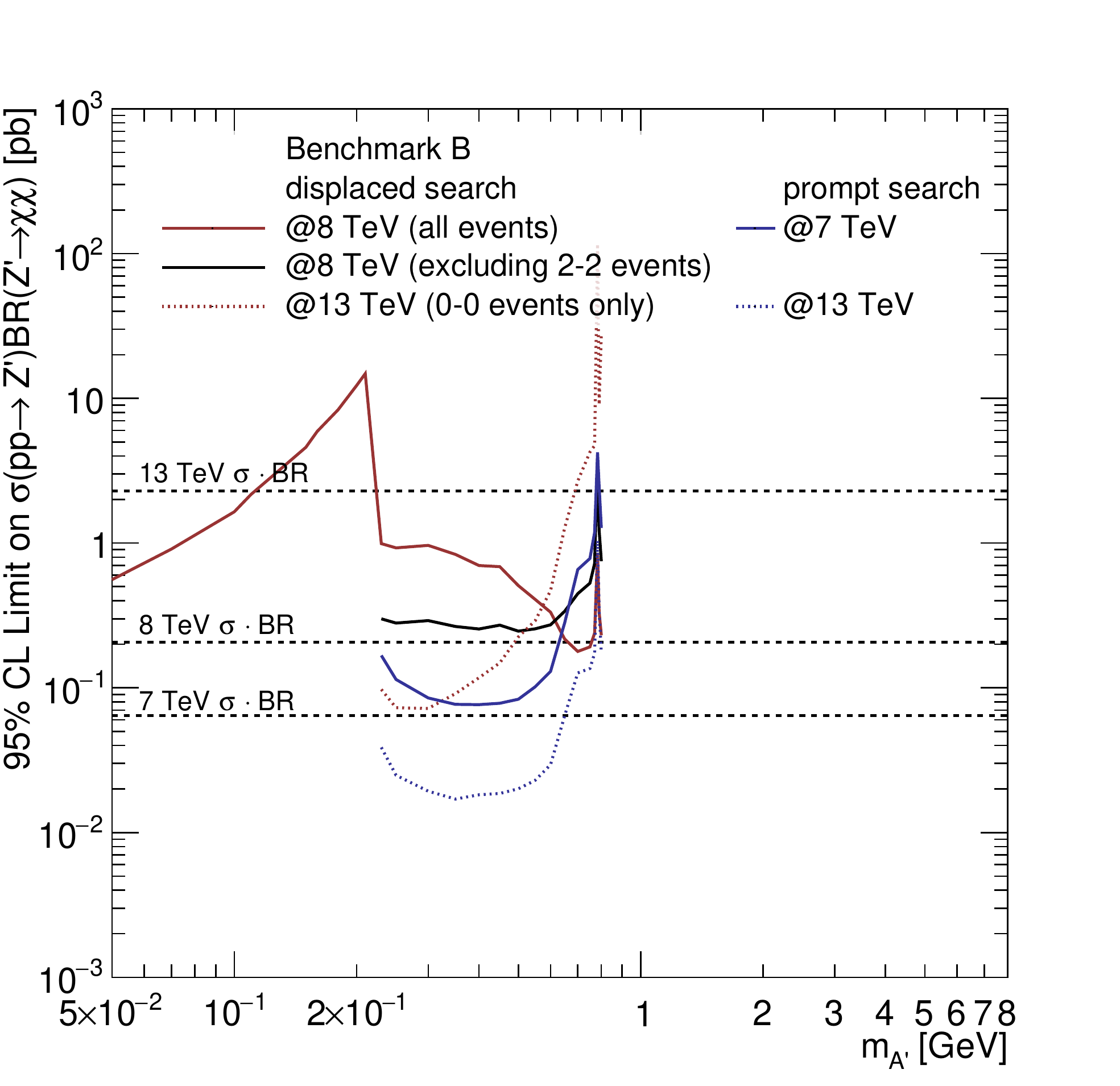} &
    \includegraphics[width=0.45\textwidth]{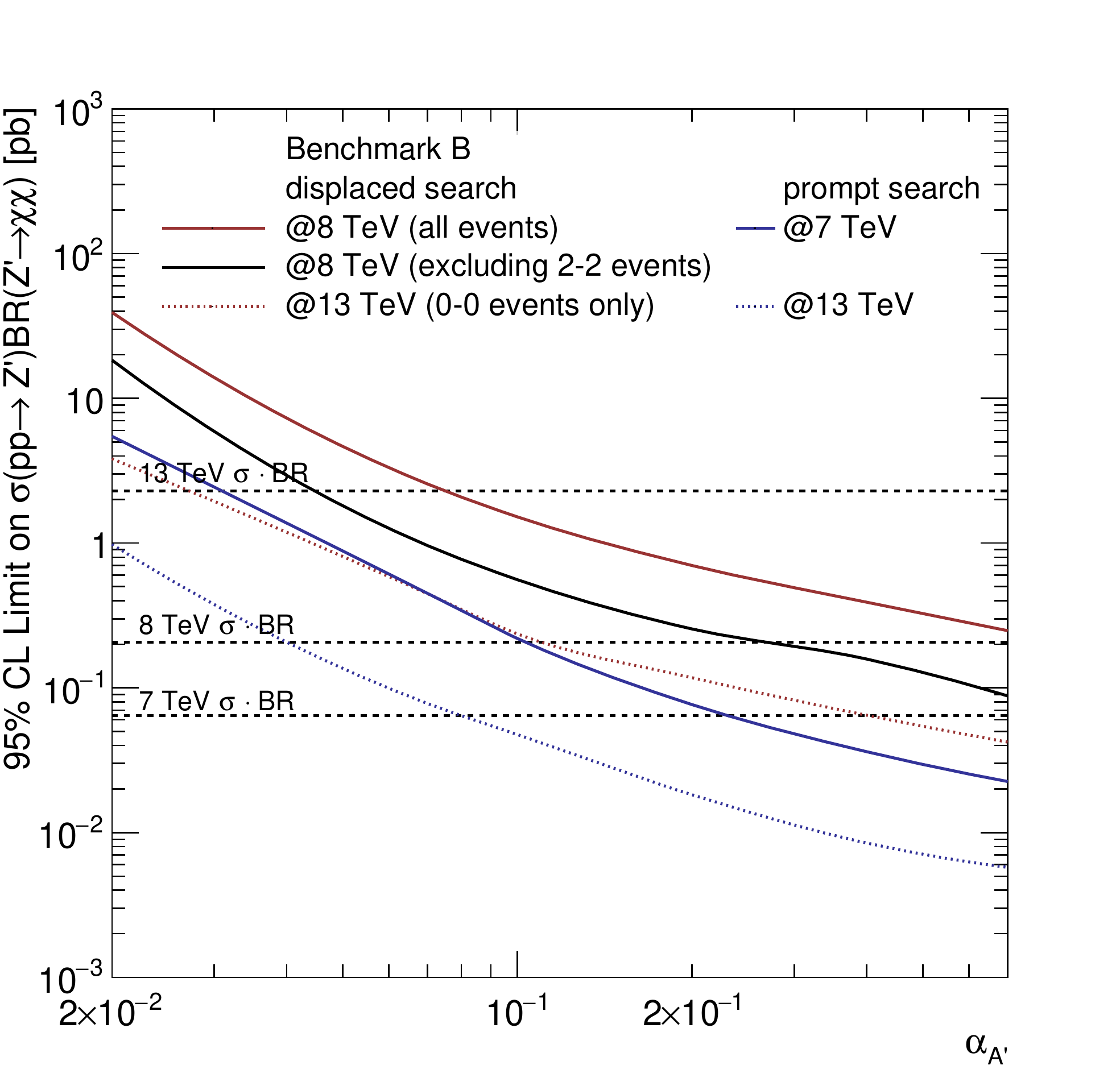} \\[-0.3cm]
    (c) & (d) \\[-0.4cm]
    \includegraphics[width=0.45\textwidth]{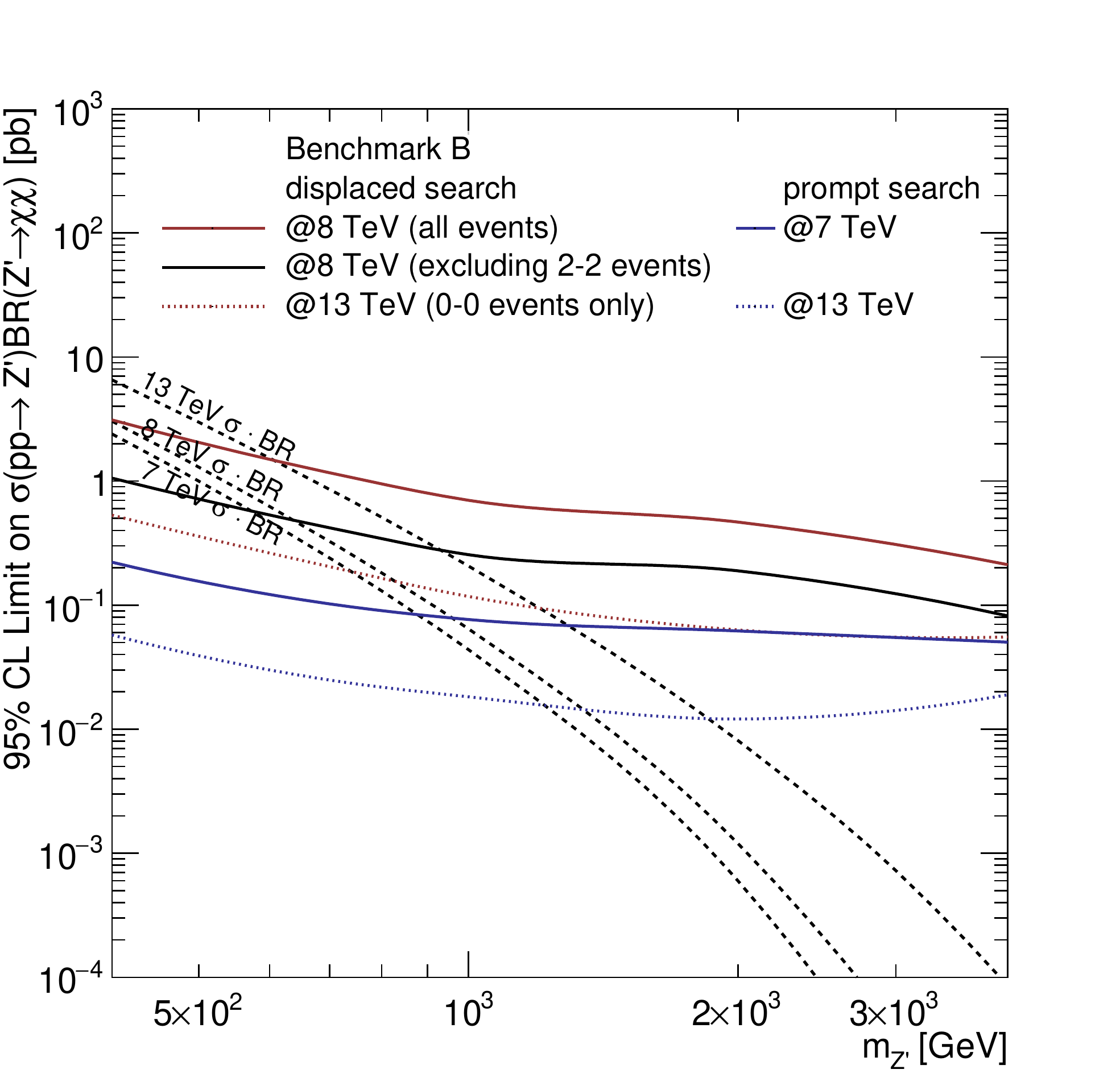} &
    \includegraphics[width=0.45\textwidth]{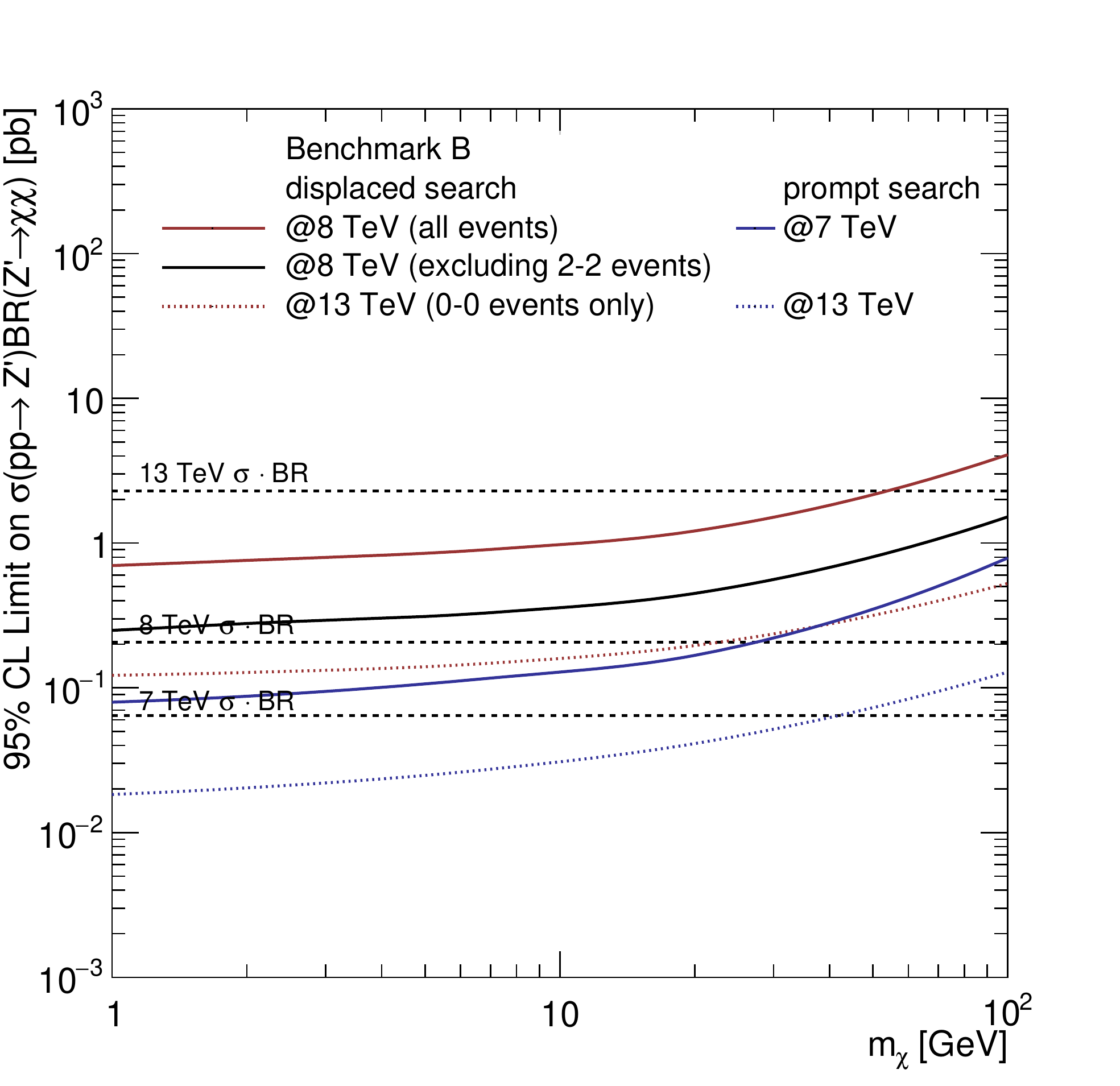} \\[-0.3cm]
    (e) & (f)
  \end{tabular}
  \end{center}
  \vspace{-0.3cm}
  \caption{Same as fig.~\ref{fig:CLs1}, but for benchmark point~B from
    table~\ref{tab:Benchmark}.}
  \label{fig:CLs2}
\end{figure}

We use the CLs method~\cite{Read:2002hq} to compare our predictions at
$\sqrt{s} = 7$~TeV and $\sqrt{s} = 8$~TeV to the data and to constrain the
production cross section for a DM pair, $\sigma(pp\to\heavymed)
\BR(\heavymed\to\bar\chi\chi)$.  We use the ATLAS background
predictions~\cite{Aad:2012qua,Aad:2014yea} as the null hypothesis, and the sum
of the ATLAS background and our signal prediction as the test hypothesis.  For
$\sqrt{s} = 13$~TeV, we compare our signal+background prediction to the
background-only prediction to obtain the expected limit.
We assume a systematic uncertainty of 30\% on the signal normalization
to account for errors in the signal and detector simulation.  The results are
presented in figs.~\ref{fig:CLs1} and \ref{fig:CLs2} for the two benchmark
points from table~\ref{tab:Benchmark}, where in each panel one of the model
parameters is varied while the others are kept constant.  For the displaced
search at $\sqrt{s} = 8$~TeV, two different limits are calculated, one
including all tagged events (solid red curves) and one excluding type 2--2
events, i.e.\ events with two type-2 lepton jets (solid black curves).  Due to
the larger background in the 2--2 sample, this second limit is stronger in
certain parameter regimes than the limit including all lepton jets.  For the
displaced search at $\sqrt{s} = 13$~TeV, (red dotted curves), we use only
events with two type-0 lepton jets, as explained above.  We also show in
figs.~\ref{fig:CLs1} and \ref{fig:CLs2} the predicted values for
$\sigma(pp\to\heavymed) \BR(\heavymed\to\bar\chi\chi)$ as thin dotted curves.

Let us first discuss fig.~\ref{fig:CLs1}, corresponding to benchmark point~A.
(We will comment on the differences compared to fig.~\ref{fig:CLs2} below.)
Since for fixed $m_\darkmed$, the dark photon decay length $c\tau$ depends only
on the kinetic mixing parameter $\epsilon$, figs.~\ref{fig:CLs1} (a) and (b)
are equivalent. We see that the best limits from the prompt search are obtained
at $c\tau \lesssim 1$~mm ($\epsilon \gtrsim 10^{-5}$), while for larger decay
length, the cut on the impact parameter $|d_0| < 1$~mm restricts the sensitivity.
For the displaced search, the sensitivity is optimal around $c\tau=10$--100~mm,
corresponding to $\epsilon \simeq \text{few} \times 10^{-6}$.  For shorter
lifetimes (larger $\epsilon$) the dark photon decay vertices tend to be closer
to the primary interaction point, so that the $\darkmed$ decay products are
more likely to leave tracks in the inner detector, thus failing the displaced
lepton jet selection criteria.  For too large lifetimes (smaller $\epsilon$),
on the other hand, most dark photons will decay outside the ATLAS detector and
not lead to a signal at all.  The exact position of the most sensitive regime
depends strongly on the $\darkmed$ kinematics and its mass.

For the displaced search, notice also the slight shift of the most sensitive
point in $c\tau$ and $\epsilon$ between the red solid, black solid and red
dotted curves in figs.~\ref{fig:CLs1} (a) and (b).  This is related mostly to
the inclusion/exclusion of different types of lepton jets in the three cases.
In particular, as explained above, we include only type-0 (muonic) lepton jets
in the 13~TeV analysis. The decay $\darkmed \to \mu^+\mu^-$ can be successfully
reconstructed for shorter $\darkmed$ decay lengths than many other $\darkmed$
decay modes (see table~\ref{tab:DetectorVolume}) Consequently, the exclusion
limit based on displaced type 0--0 events only (red dotted curves in
fig.~\ref{fig:CLs1} (a) in (b)) is best at relatively small $c\tau$ (relatively
large $\epsilon$), followed by the sample excluding type 2--2 events and by the
sample including all types of lepton jets.

Similar arguments help us understand the exclusion limits as a function of the
$\darkmed$ mass in fig.~\ref{fig:CLs1} (c). The different peaks and dips in
this plot reflect the dependence of the $\darkmed$ decay branching ratios on
$m_\darkmed$ illustrated in fig.~\ref{fig:BR}, see also \cite{Agashe:2014kda}.
The most notable features arise
due to the broad $\rho^0$ resonance around 780~MeV with the narrow $\omega$
resonance on top of it and the narrow $\phi$ resonance at 1~GeV.  The $\rho^0$
resonance decays almost exclusively to $\pi^+ \pi^-$ pairs, so that around this
resonance, the relative importance of all other decay channels is diminished.
This implies that the prompt lepton jet search, which includes only muonic
lepton jets, has a greatly reduced sensitivity around the $\rho$ resonance.  In
the displaced search, most $\darkmed \to \rho^0 \to \pi^+\pi^-$ decays are
reconstructed as type~2 (calorimeter) lepton jets (see
table~\ref{tab:DetectorVolume}), therefore the sensitivity improves around the
$\rho^0$ resonance if type~2--2 events are included in the analysis (red
solid curve in fig.~\ref{fig:CLs1} (c)). If type 2--2 events are excluded (black
solid and red dotted curves in fig.~\ref{fig:CLs1} (c)), the sensitivity
decreases.  The dominant decay of the $\omega$ resonance is to $\pi^+ \pi^-
\pi^0$.  This again decreases the sensitivity of the prompt search around the
resonance. For the displaced search, we can read from
table~\ref{tab:DetectorVolume} that the decay $\darkmed \to \omega \to \pi^+
\pi^- \pi^0$ is more likely to be vetoed than $\darkmed \to \pi^+ \pi^-$, therefore
the sensitivity is reduced at the $\omega$ resonance compared to the
off-resonance region.  The behavior at the $\phi$ resonance is similar to that
at the $\rho^0$ resonance. The prompt search is insensitive because the $\phi$
mostly decays hadronically.  For the displaced search, detection prospects for
the dominant decay channel $\phi \to K^+K^-$ (branching ratio 48.9\%) as a
type-2 (calorimeter) lepton jet are very similar to those of the $\rho^0$
resonance decaying to $\pi^+ \pi^-$ (see table~\ref{tab:DetectorVolume}).  The
secondary decay $\phi \to K^0_L K^0_S$ (branching ratio 34.2\%)
has a high probability of being
reconstructed as a displaced type-2 lepton jet even if the $\darkmed$ decay
happens in the inner detector.  For these reasons, the sensitivity at the
$\phi$ resonance is further boosted in analyses including displaced type-2
lepton jets. For $\darkmed$ masses above 1~GeV, the sensitivities of both the
prompt and the displaced searches decrease rapidly. For the prompt search, the
invariant mass requirement $m_{\mu\mu} < 2$~GeV restricts the realm of
sensitivity to $m_\darkmed < 2$~GeV.  For the displaced search, the smaller
$\darkmed$ boost factor at larger $m_\darkmed$ implies that the $\darkmed$
decay length becomes shorter and events are more likely to leave tracks in the
inner detector and be vetoed. Moreover, the average number of radiated
$\darkmed$ decreases at larger $m_\darkmed$.  As discussed in
sec.~\ref{sec:model}, we do not consider the region $1.7\ \text{GeV} <
m_\darkmed < 2.3\ \text{GeV}$, where numerous hadronic resonances appear but
the quark model is not applicable yet. In fig.\ref{fig:CLs1} (c), we have
smoothly interpolated through this region. As can be seen, the low-energy
description in terms of hadron final states and the high-energy description in
terms of quark final states match onto each other very well.

In fig.~\ref{fig:CLs1} (d), we find that, unsurprisingly, all limits
improve as the dark sector gauge coupling $\alpha_\darkmed$ increases
because the probability for $\darkmed$ radiation grows. Note, however,
that in the very large $\alpha_\darkmed$ region, our perturbative treatment
of the dark photon shower is expected to break down.

Concerning the dependence of the sensitivity on the $\heavymed$ mass
(fig.~\ref{fig:CLs2} (e)), we find that at $m_\heavymed < \text{few TeV}$, the
sensitivity of all searches increases with $m_\heavymed$. This is because a
heavier $\heavymed$ decays to more energetic $\chi$ particles, which radiate
more dark photons. Eventually, of course, on-shell production of the $\heavymed$
becomes impossible and the sensitivity decreases again.

Finally, fig.~\ref{fig:CLs1} (f) shows that the sensitivity goes down when
$\chi$ is made heavier because less boosted $\chi$ particles radiate less.

Let us now turn to a comparison of the two benchmark models from
table~\ref{tab:Benchmark} and compare figs.~\ref{fig:CLs1} and \ref{fig:CLs2}.
The general features of the two figures are very similar, but in general
benchmark point~B is more easily detectable in prompt lepton jet searches due to
the smaller $c\tau$.  Benchmark point~B offers somewhat better sensitivity than
benchmark point~A also because the smaller values of $m_\chi$ and $m_\darkmed$
imply that the average number of $\darkmed$ radiated in each event is larger.
This effect is especially pronounced in the displaced searches excluding type
2--2 events, which are limited by backgrounds.  Note that in
fig.~\ref{fig:CLs2} (c) the region with $m_\darkmed > 2 m_\chi$ is not
considered because it would lead to a very large branching ratio for $\darkmed
\to \bar\chi \chi$, making the $\darkmed$ invisible to the detector.

\begin{figure}
  \begin{center}
    \begin{tabular}{cc}
      \includegraphics[width=0.48\textwidth]{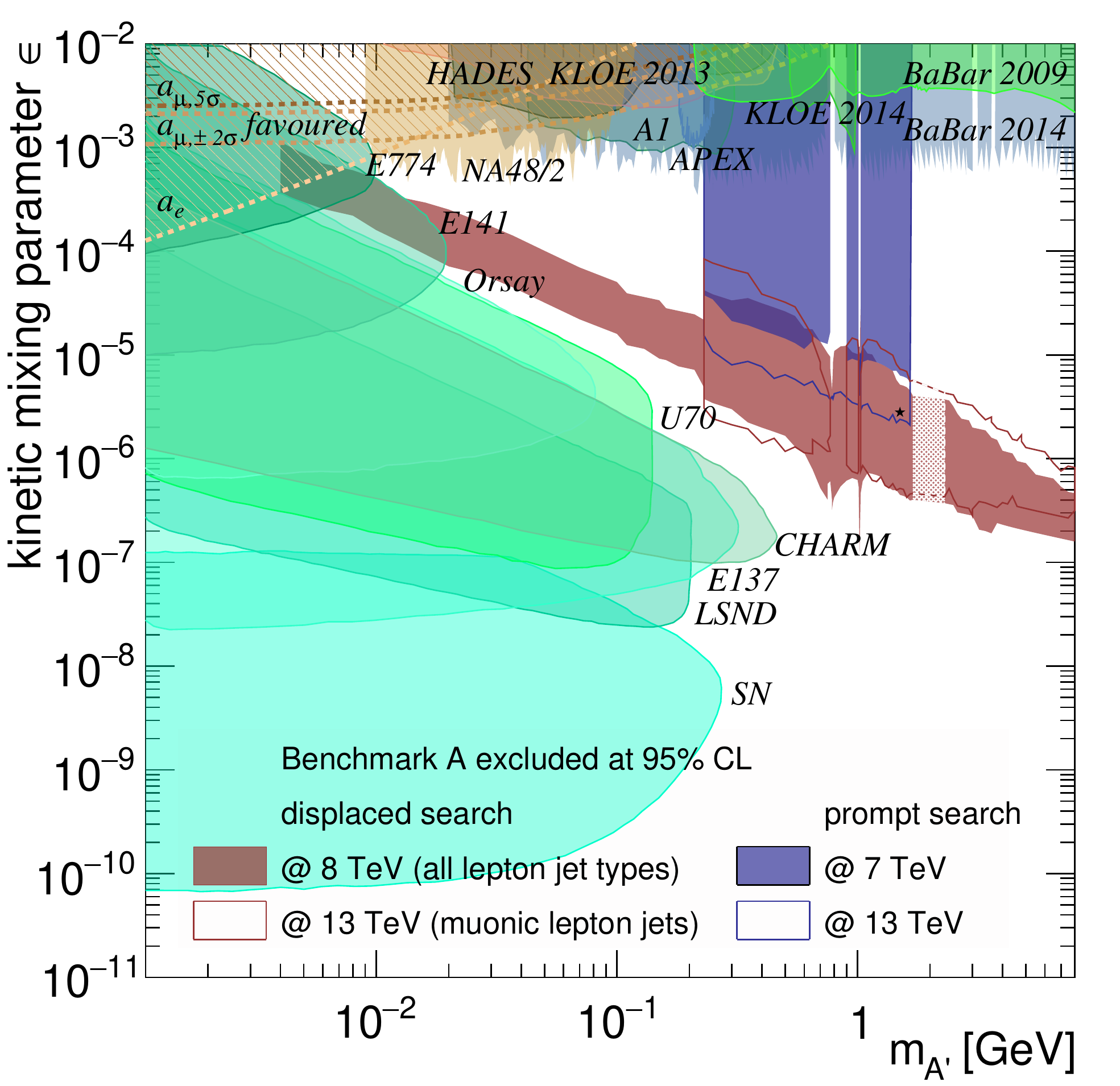} &
      \includegraphics[width=0.48\textwidth]{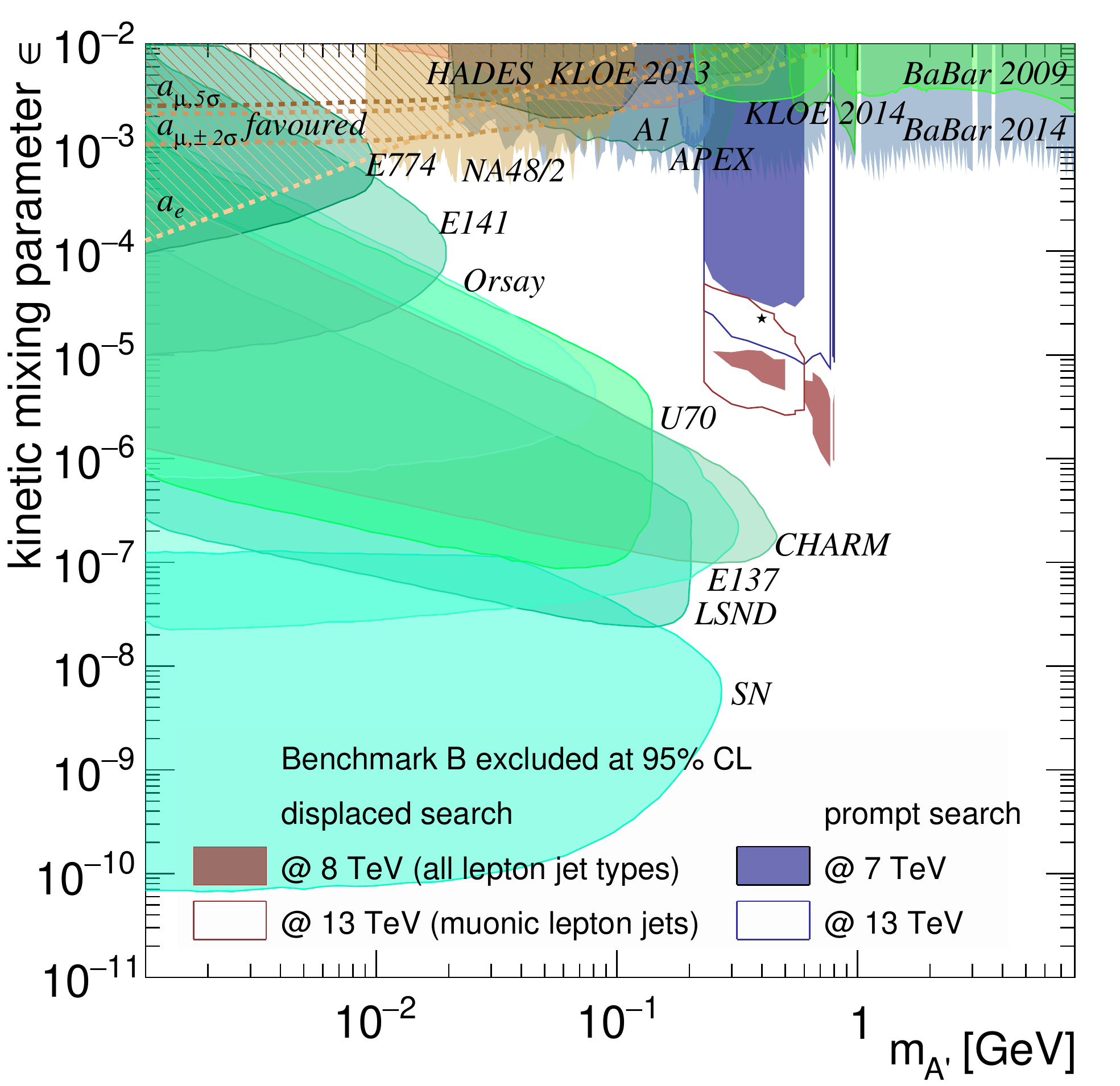} \\
      (a) & (b)
    \end{tabular}
  \end{center}
  \caption{95\% CL constraints on the dark photon mass $m_\darkmed$ and the kinetic
    mixing parameter $\epsilon$, with all other model parameters fixed at the
    first (second) set of benchmark values from table~\ref{tab:Benchmark} in
    the left hand (right hand) panel.  We show exclusion limits from the ATLAS
    search for prompt lepton jets in 5~fb$^{-1}$ of 7~TeV
    data~\cite{Aad:2012qua} (blue shaded region) and from their displaced
    lepton jet search in 20.3~fb$^{-1}$ of 8~TeV data~\cite{Aad:2014yea} (red
    shaded region), as well as projected sensitivities for 100~fb$^{-1}$ of
    13~TeV data (blue/red unshaded regions). Black stars correspond to the nominal
    values of $m_\darkmed$ and $\epsilon$ from table~\ref{tab:Benchmark}.
    The lighter colored region around
    $m_\darkmed=2$ GeV corresponds to the transition region between the
    analysis in terms of hadron final states and the analysis in terms of quark
    final states and is based on interpolation.  We also show the existing 90\%
    CL exclusion limits from the electron and muon anomalous magnetic
    moment~\cite{Pospelov:2008zw,Davoudiasl:2012ig,Endo:2012hp},
    HADES~\cite{Agakishiev:2013fwl}, KLOE 2013~\cite{Babusci:2012cr} and
    2014~\cite{Babusci:2014sta}, the test run results from
    APEX~\cite{Abrahamyan:2011gv}, BaBar 2009~\cite{Aubert:2009cp} and
    2014~\cite{Lees:2014xha}, beam dump experiments E137, E141, and
    E774~\cite{Blumlein:2011mv,Bjorken:2009mm,Bross:1989mp},
    A1~\cite{Merkel:2011ze}, Orsay~\cite{Davier:1989wz},
    U70~\cite{Blumlein:2013cua}, CHARM~\cite{Gninenko:2012eq},
    LSND~\cite{Essig:2010gu}, as well as constraints from astrophysical
    observations~\cite{Dent:2012mx,Dreiner:2013mua} and $\pi^0$
    decays~\cite{CERNNA48/2:2015lha}.}
  \label{fig:CLs2D}
\end{figure}

To put our results in the context of other constraints on dark photons, we show
in fig.~\ref{fig:CLs2D} the dark photon parameter space spanned by $\epsilon$
and $m_\darkmed$.  For both of our benchmark points, we compare the limits we
derived from the prompt and displaced ATLAS lepton jet searches (blue
shaded/red shaded) and the predictions for 13~TeV (blue/red unshaded) to the
exclusion limits from various low energy experiments.  For the 8~TeV displaced
analysis, we decide for each parameter space point individually whether or not
to exclude type 2--2 events (events with two calorimeter lepton jets),
depending on which analysis leads to the better expected limit.

We see that the parameter region probed by LHC lepton jet searches is
complementary to the region probed by low energy searches. It is important to
keep in mind, though, that low energy experiments probe any dark dark photon
model, while our analysis is sensitive only to scenarios that, in addition,
feature a light DM particle that can be pair-produced at the LHC.  We see that
benchmark point~A is already excluded by the 8~TeV ATLAS search for displaced
lepton jets, while benchmark point~B can be probed by both the prompt and the
displaced search at 13~TeV.  The exclusion regions from the displaced searches
move to smaller values of $\epsilon$ at larger $m_\darkmed$ because the
$\darkmed$ width increases
and the lab frame decay length decreases with increasing $m_\darkmed$ due
to the smaller boost factors. Both effects need to be compensated by a
decrease in $\epsilon$ to avoid
$\darkmed$ decays in the inner detector.  The impact of the various $\darkmed$ decay channels
on the sensitivity is again reflected by spikes, dips and kinks in the excluded
regions. For instance, note that the prompt search at 7~TeV, which includes
only muonic lepton jets, is insensitive close to the hadronically decaying
$\rho$, $\omega$ and $\phi$ resonances. At 13~TeV, the gap is closed
thanks to the much better statistics.
The sharp edge at low $\darkmed$ mass for the 7~TeV and 13~TeV regions
corresponds to the opening of the decay channel $\darkmed \to \mu^+ \mu^-$ at
$m_\darkmed = 2 m_\mu$. Remember that, in our 7~TeV and 13~TeV analysis, we
could only include muonic lepton jets, therefore, our analysis is insensitive
below the muon threshold. A full experimental search would of course extend
also into this region, like the 8~TeV ATLAS search for displaced lepton jets
has done.

\section{Conclusions}
\label{sec:conclusions}

In summary, we have studied the possibility of revealing the properties of a
dark sector of particle physics by using final state radiation from dark matter
produced at the LHC.  The characteristic experimental signature of this process
is a pair of lepton jets.  While our conclusions apply to a very large class of
extended dark sector models, we have worked in the framework of a toy model
where fermionic dark matter particles $\chi$ are charged under a new dark
sector gauge group $U(1)'$ and coupled to the SM through a heavy mediator.
When $\chi$ particles are pair-produced at the LHC via this heavy mediator,
they can radiate several light $U(1)'$ gauge bosons $\darkmed$ (dark photons)
which subsequently decay to light SM particles (electrons, muons, mesons)
through a small kinetic mixing $\epsilon$ with the SM photon.  Due to the
required smallness of the kinetic mixing, the $\darkmed$ decay length can be
macroscopic.  We emphasize that our results for this toy model are easily
generalized to any model with light dark sector particles charged under a new
gauge interaction.  We also remind the reader that, in order to account for all
of the DM in the Universe, $\chi$ must be produced non-thermally, as for
instance in asymmetric DM scenarios.

We have first developed two analytic treatments of dark photon radiation: in
the first one, we use recursive expressions for the $\darkmed$ and $\chi$
energy distributions, while in the second one, we compute the moments of these
distributions fully analytically and then apply an inverse Mellin transform to
obtain the distributions themselves.  We have compared our analytic
calculations with Monte Carlo simulations in Pythia, finding excellent
agreement.

In the second part of the paper, we have extended these Monte Carlo simulations
by a simplified description of the ATLAS detector that allows us to recast the
existing ATLAS searches for prompt and displaced lepton jets into powerful
limits on the DM pair production cross section and the dark photon parameters
in radiating DM models.  Our limits on the $\bar\chi \chi$ production cross
section range down to $\mathcal{O}(10\,\text{fb})$, depending on the $\darkmed$ mass and
lifetime, the $U(1)'$ gauge coupling, and the mass of $\chi$.  Regarding the
$\darkmed$ properties, we find that LHC searches for radiating DM can probe a
region in $m_\darkmed$--$\epsilon$ space that has been inaccessible to low
energy dark photon searches so far, namely in the parameter range $10^{-7}
\lesssim \epsilon \lesssim 10^{-3}$ for $m_\darkmed$ in the MeV--GeV range. Of
course, these limits are subject to the condition that DM particles with large
coupling to dark photons ($\alpha_\darkmed \sim \mathcal{O}(0.1)$) can be
pair-produced at the LHC in significant numbers.

Looking into the future, we have also shown that LHC limits will significantly
improve in Run~2 at 13~TeV center of mass energy. It is worth emphasizing here
that our simplified 13~TeV analyses are still a far cry from what a full
experimental search can achieve. Most importantly, with a more detailed
detector simulation and data-driven background estimation methods, it will be
possible to include not only muonic lepton jets, but also $\darkmed$ decays
to electrons and hadrons.

To conclude, we have shown that lepton jets are an interesting and powerful
tool to elucidate the dynamics of the dark matter sector. Together with searches
for hadronic jets with non-standard properties, they will be essential in
the hunt for dark matter at Run~2 of the LHC and might well be the first
to find a positive signal.

\begin{acknowledgments}
  It is a pleasure to thank Guido Ciapetti, Stefano Giagu, Antonio Policicchio
  and Christian Schmitt for very helpful discussions on the ATLAS detector and
  the ATLAS lepton jet analyses.  We are also grateful to Torbj\"{o}rn
  Sj\"{o}strand for clarifying some of the subtleties of Pythia for us, and to
  Tilman Plehn for useful conversations in the very early stages of this
  project.  PANM would like to thank the Mainz Institute for Theoretical
  Physics (MITP) for support during two visits to Mainz that were crucial for
  this project.  JK and JL are supported by the German Research Foundation
  (DFG) under Grant No.\ \mbox{KO~4820/1--1}.  PANM acknowledges partial
  support from the European Union FP7 ITN INVISIBLES (Marie Curie Actions,
  PITN-GA-2011-289442) and from the Spanish MINECO's ``Centro de Excelencia
  Severo Ochoa'' Programme under grant SEV-2012-0249.
\end{acknowledgments}

\bibliographystyle{JHEP}
\bibliography{refs-DR}

\end{document}